\documentclass[letterpaper,12pt]{article}

\usepackage{graphicx}                % for importing graphics into figures
\usepackage{float}                   % better control of figure placement
\usepackage{hyperref}                % for clickable URLs and email addresses
\usepackage[margin=1in]{geometry}  % control page margins
\usepackage[short]{datetime}         % precise date/time stamp on titlepage
\usepackage[labelfont=bf]{caption}   % make caption labels boldface
\usepackage[bottom]{footmisc}        % footnotes at bottom of page
\usepackage{lineno}
\usepackage{ragged2e}
\usepackage[protrusion=true,expansion=true]{microtype} % Better typography
\usepackage[T1]{fontenc}                               % Better typography
\usepackage{lmodern}
\usepackage{tikz}

\usepackage[compact]{titlesec}
\titleformat*{\section}{\large\bfseries}
\titlespacing*{\section}{0pt}{1ex}{1ex}
\titleformat*{\subsection}{\normalsize\bfseries}
\titlespacing*{\subsection}{0pt}{0ex}{0ex}
\titleformat*{\subsubsection}{\normalsize\bfseries}
\titlespacing*{\subsubsection}{0pt}{0ex}{0ex}

\usepackage{amsfonts,amssymb,amsmath, mathtools}
\usepackage{xcolor, color, soul}
\usepackage{graphicx}
\usepackage{multicol}
\usepackage{url}
\usepackage{faktor}
\usepackage{ulem}
\usepackage{pdfpages}
\usepackage{soul}
\usepackage{hyperref}
\usepackage{bm}
\usepackage{natbib}
\usepackage{subcaption}
\usepackage{booktabs}
\usepackage{setspace}
\usepackage{amsthm}

\def\indep{{\,\perp \!\!\! \perp\,}}

\newcommand{\Var}[0]{\text{Var}}

\newtheorem{proposition}{Proposition}

\begin{document}

\center{\textbf{\Large A Unified Framework for Causal Estimand Selection}}
\vspace{-.05in} 
\center{ Martha Barnard$^{*}$, Jared D. Huling, Julian Wolfson}
\vspace{-.1in}
\center{\footnotesize Division of Biostatistics and Health Data Science, School of Public Health, University of Minnesota}
\vspace{-.1in}
\justifying
\begin{abstract}
    Estimating the causal effect of a treatment or health policy with observational data can be challenging due to an imbalance of and a lack of overlap between treated and control covariate distributions. In the presence of limited overlap, researchers choose between 1) methods (e.g., inverse probability weighting) that imply traditional estimands but whose estimators are at risk of considerable bias and variance; and 2) methods (e.g., overlap weighting) which imply a different estimand, thereby modifying the target population to reduce variance. We propose a framework for navigating the tradeoffs between variance and bias due to imbalance and lack of overlap and the targeting of the estimand of scientific interest. We introduce a bias decomposition that encapsulates bias due to 1) the statistical bias of the estimator; and 2) estimand mismatch, i.e., deviation from the population of interest. We propose two design-based metrics and an estimand selection procedure that help illustrate the tradeoffs between these sources of bias and variance of the resulting estimators. Our procedure allows analysts to incorporate their domain-specific preference for preservation of the original research population versus reduction of statistical bias. We demonstrate how to select an estimand based on these preferences with an application to right heart catheterization data.
\end{abstract}

%  Please place your key words in alphabetical order, separated
%  by semicolons, with the first letter of the first word capitalized,
%  and a period at the end of the list.
%
\vspace{-0.1in}
\noindent\rule{2in}{0.4pt} \\
{\small * barna126@umn.edu}

\thispagestyle{empty}

\clearpage
\setcounter{page}{1}

\section{Introduction}

%A primary goal of many observational studies is to determine the causal effect of a treatment on an outcome. 
A primary goal of healthcare research is to determine the causal effect of a treatment or policy on health outcomes. Estimating causal effects using observational data requires 1) an unconfounded comparison of treated and control groups such that the two groups have similar, or balanced, characteristics; and 2) overlap between the treated and control groups such that all observations have a non-zero probability of being assigned either treatment or control. 
%When covariate distributions are imbalanced, treatment effect estimates can have bias because the treated and control groups are not comparable. When there is a lack of overlap, estimators have the potential for both bias and variance inflation because the treatment group data fails to represent the entire population of interest. 
While there are a wide variety of methods to adjust for covariate imbalance, overlap is less well-understood and there are relatively fewer methods that address its impacts on the estimation of causal effects. 
Some methods aim to reduce variance inflation caused by limited overlap through either achieving only approximate balance or by modifying the population of interest to a new population has sufficient overlap. Yet, these approaches, respectively, increase estimator bias or shift the target estimand away from the original estimand of scientific interest. It is also often unclear which method, or which estimand, is best suited for a given scientific question and degree of overlap. In this paper, we develop a framework for  simultaneously navigating the tradeoffs between preserving the original target population and minimizing the bias and variance that can result from imbalance and lack of overlap.

\subsection{Weighting methods for balancing covariates}
There are many methods that balance the covariates between groups (e.g., regression, matching, weighting). We focus on weighting methods which reweight the data in order to achieve covariate balance. One set of such methods seek to directly balance the covariates by deriving weights subject to covariate constraints using optimization procedures \citep{hainmueller_entropy_2012, zubizarreta_stable_2015, wang_minimal_2020}. Another set of methods rely on inverse probability weighting (IPW), where observations are inverse weighted by the sample probability of receiving their assigned treatment \citep{rosenbaum_assessing_1983b,hahn_role_1998,robins_marginal_2000,hirano2001,hirano_efficient_2003,imbens_nonparametric_2004}. Both types of methods can target estimands corresponding to a specific population, such as the average treatment effect defined over the entire (ATE), treated (ATT), or control (ATC) sample population. While these methods target estimands with a meaningful population, a lack of overlap between the treated and control covariate distributions can create very large weights that produce estimators with inflated variance and the potential for considerable bias \citep{busso_new_2014, hong_inference_2020}. While direct balancing methods generally provide more stable and smaller weights, a lack of overlap can yield no solutions to the optimization problem \citep{hainmueller_entropy_2012, chattopadhyay_balancing_2020}. Similar issues arise with matching methods, where large differences between the treated and control groups can result in bias in the treatment effect estimator due to imbalance.

\subsection{Weighting methods that address lack of overlap}
A variety of new weighting methods have been created that mitigate the impacts of lack of overlap on estimation. For the majority of these methods, reducing variance of the estimator is the primary goal. 
%Within this set of methods there are two types: 1) estimator variance is reduced through increasing estimator bias; and 2) estimator variance is reduced through changing the estimand of interest such that the corresponding population has better overlap. Within this first type of method, 
\cite{wang_minimal_2020} explore a class of direct balancing weights called minimal weights which provide approximate, rather than exact balance of the covariate distributions, resulting in estimators with reduced variance but potentially increased bias.
%increases the bias of the estimator while reducing the variance. 
%While estimator bias may be increased with this method, the estimand and therefore the target population remains unchanged. 
Other approaches work by modifying
%Most new methods fall into the second category of methods that modify 
the estimand, and consequently the target population, either explicitly or implicitly to achieve better overlap. \cite{crump_dealing_2009} formalize modifying the estimand through propensity score trimming while \cite{li_weighting_2013} and \cite{yang_asymptotic_2018} propose continuous weights that approximate pair matching and propensity score trimming, respectively. One of the more widely-used such methods is overlap weights \citep{li2018, li2019}, which
minimize the asymptotic variance of the weighted average treatment effect estimator across
all estimands characterized by weighted populations under the assumption of homoscedasticity.
%which weight each individual by the probability of being assigned to the opposite group.
Overlap weights emphasize individuals with propensity scores around $0.5$, i.e. those with so-called clinical equipoise, targeting the ``overlap'' (ATO) population. 
%When the variance of outcomes is constant, overlap weights minimize the asymptotic variance of the weighted average treatment effect estimator across all possible estimands characterized by weighted populations. 
%Simulation studies have shown that overlap weights perform well under model misspecification and can improve statistical power compared to IPW weights \citep{zhou_propensity_2020,mao2019}. 
Matching methods also often address lack of overlap by modifying the matched population and frequently operate by discarding hard-to-match samples, improving balance by modifying the population  \citep{rosenbaum_optimal_2012, de_los_angeles_resa_evaluation_2016,visconti_handling_2018}. 
%For all these methods, the estimand of interest, and therefore the population, is changed in order to achieve better statistical performance of the estimator.

There are a variety of scenarios in which it is reasonable to deviate from estimands that aggregate over the entire sample population; for example, some sample populations are a convenience sample 
%rather than a meaningful population 
\citep{rosenbaum_optimal_2012} and as such, the ATE %over such a population 
may not have clinical relevance. 
%However, in many other scenarios the sample population or treated population is meaningful, making it still desirable to estimate the ATE or ATT even in the presence of low overlap.
\cite{li2018} and \cite{mao2019} both argue that overlap weights target a population of interest; people with propensity scores around $0.5$ may be the individuals for whom clinical treatment guidelines are unclear. 
%or the individuals who would be more receptive to policy shifts.
While this is often true, in practice treatments and policies may not be assigned equitably along demographic factors (\cite{faigle_racial_2017,raeisi-giglou_disparities_2022, tchikrizov_health_2023}); in these cases, the interpretation of individuals at equipoise is unclear. 
%in these cases, it is unclear whether individuals at equipoise in the data truly reflect a group of people for whom clinical guidelines are unknown. 
Regardless of whether clinical equipoise is of interest, it can be hard to describe the exact population that is induced by overlap weights. Therefore, in many scenarios it is still desirable to estimate the ATE or ATT even in the presence of low overlap to preserve a meaningful and/or interpretable population. Among methods that address lack of overlap, there has been limited development on constructing an estimand that also has an interpretable population. \cite{traskin_defining_2011} and \cite{fogarty_discrete_2016} propose methods for selecting an interpretable population with better overlap, %through tree and discrete optimization techniques, respectively,
though both methods can still produce estimators with considerable bias and/or variance.

%Cardinality matching provides an option for constraining the matched set to be similar to some population of interest \citep{visconti_handling_2018}. 
%\cite{traskin_defining_2011} and \cite{fogarty_discrete_2016} propose methods for selecting an interpretable population with better overlap, through tree and discrete optimization techniques, respectively, though both methods can still produce populations that yield estimators with considerable bias and/or variance.

\subsection{Proposed framework}

There are clear tradeoffs between retaining an estimand with a meaningful target population, estimator bias, and variance among current methods for treatment effect estimation when there is a lack of overlap. Standard weighting methods target estimands with meaningful populations, however there is risk of estimator bias and inflated variance. Methods such as minimal weights and overlap weights aim to %either derive a set of weights or identify an estimand that 
minimizes estimator variance, but in doing so they either sacrifice estimator bias or eschew the original target population. Most methods described above aim to improve or maintain only one of estimator bias, variance, or the target population, but the extent to which the other two characteristics are influenced for a given analysis is often unclear. 
%% commenting out because it's sort of repeated two sentences later:
%; there are few, if any, tools for the quantifying the extent to which the other two characteristics are influenced for a given analysis. 
Different biomedical or healthcare application may seek to prioritize these three characteristics differently; 
in some cases statistical performance may be the primary priority, while in others it may be important to maintain the population of interest. 
%In some scenarios, even if the target estimand is the ATE, the variance reductions when targeting a slightly different estimand may outweigh the risk of bias due to mismatch from the original target population. 
However, as far as we know, there are few or no methods that provide tools to flexibly explore and quantify the tradeoffs between all three characteristics. 

In this work, we present a conceptual framework and corresponding procedure for characterizing and selecting an estimand according to its target population, estimator bias, and variance, allowing the analyst to explicitly prioritize any of these properties. As part of the conceptual framework, we introduce a bias decomposition that encapsulates bias due to 1) the statistical bias of the estimator; and 2) estimand mismatch i.e., deviation from the population of interest. To leverage this decomposition for estimand selection, we propose two design-based metrics that target each part of the bias decomposition. Furthermore, we propose an estimand selection procedure that enumerates a sequence of estimands corresponding to different tradeoffs between preservation of the target estimand and statistical performance in terms of bias and variance, where the ATE, ATT, ATC, and ATO are special cases of the set of possible estimands under consideration. %To elicit this set of estimands, we propose two design-based metrics that target each part of the bias decomposition. Furthermore, we propose an estimand selection procedure that clearly illustrates the bias-variance tradeoffs of potential estimands. 
We analyze synthetic data from a variety of data generating scenarios and data from a right heart catheterization study to demonstrate the effectiveness of our method in navigating these tradeoffs. We present all methods and results in reference to the ATE as the estimand of scientific interest, however, all methods can apply to other estimands/populations such as the ATT. 

The remainder of the paper is outlined as follows. In Section \ref{sec:background}, we introduce notation and our conceptual framework for characterizing and selecting estimands. Section \ref{sec:methods} describes the design-based metrics for characterizing both sources bias and the proposed estimand selection procedure. Section \ref{sec:simulation} validates our metrics and evaluates the performance of our estimation selection procedure in comparison to the ATE and ATO through simulation studies. In Section \ref{sec:real_data}  we apply our entire procedure to reanalyze a study on right heart catheterization. Finally, Section \ref{sec:discussion} discusses the results and future work. 

\section{Conceptual framework for causal estimand selection}
\label{sec:background}
\subsection{Notation and assumptions}
Consider a sample $\{(Z_i, Y_i, \bm{X}_i)\}_{i=1}^n$ of size $n$ from a population. For individual or unit $i$, let $Z_i = z$, indicate belonging to the treatment ($z=1$) or control group ($z=0$), $Y_i$ be the outcome, and $\bm{X}_i = (X_{i1}, \ldots X_{ip})$ be a $p$ length vector of covariates. Let $e(\bm{x}) = \Pr(Z_i =1 | \bm{X}_i = \bm{x})$ be the propensity score. We use the potential outcomes framework (\cite{neyman_application_1990, rubin_estimating_1974, rubin_bayesian_1978,hernan_causal_2024}), where $Y(z)$ is the outcome that would be observed if assigned to treatment group $z$. In practice, only one potential outcome is observed per individual. We assume the standard stable unit treatment value assumption (SUTVA) that states the potential outcomes for each individual are unaffected by the treatment assignments of other individuals and that there is only one version of treatment, which implies that $Y_i = Y_i(Z_i)$. We also assume that the treatment effect is strongly unconfounded such that $\{Y(0), Y(1)\} \indep Z | \bm{X}$, which requires that there are no unmeasured confounders. Finally, we assume positivity (\cite{rosenbaum_assessing_1983b}), which requires that $0 < e(\bm{x}) < 1$ such that all individuals have a non-zero probability of being assigned to either the treatment or control group. Given these assumptions and using potential outcomes notation, the conditional average treatment effect (CATE) is $\tau(\bm{x}) = \mu_1(\bm{x}) - \mu_0(\bm{x}) = E\{Y(1) - Y(0)|\bm{X} = \bm{x}\}$. 
%Often, we desire to compute the ATE not with respect to a single $\bm{x}$ but over a specific target population or covariate distribution. 
Let $f(\bm{x})$ be the marginal distribution of the covariates and $h(\bm{x})$ some function such that $f(\bm{x})h(\bm{x})$ is the density of a target population. Then the estimand, $\tau_h$, averages the CATE over the target population,
%\vspace{-0.2in}
%\begin{align*}
  $\tau_h = \dfrac{\int \tau(\bm{x})h(\bm{x})f(\bm{x})d\bm{x}}{\int h(\bm{x})f(\bm{x})d\bm{x}}.$
%\end{align*}
 %Note that each $h(\bm{x})$ defines a different target population and thus a different causal estimand, $\tau_h$. Throughout, we assume that $\tau(\bm{x})$ can vary with $\bm{x}$, such that averaging over different populations (i.e., different causal estimands) may yield different results.
 Each $h(\bm{x})$ defines a different target population and thus a different causal estimand. We assume that $\tau(\bm{x})$ varies with $\bm{x}$, such that averaging over different populations (i.e., different causal estimands) may yield different results.
 For a given weight function $h(\bm{x})$, \cite{li2018} propose a class of propensity score balancing weights such that weighted covariate distributions are balanced between the two treatment groups,
 %\vspace{-0.3in}
\begin{align}
\label{eq:w1}
    w_1(\bm{x}) &= \frac{h(\bm{x})}{e(\bm{x})} \:\:\: \text{for} \:\:\: Z = 1, \text{ and } \:\:\:  w_0(\bm{x}) = \frac{h(\bm{x})}{1 - e(\bm{x})} \:\:\: \text{for} \:\:\: Z = 0. 
\end{align}
For the ATE, which averages  $\tau(\bm{x})$ over the entire covariate population, $h(\bm{x}) = 1$ and the balancing weights are the standard IPW weights. Other common causal estimands, such as the ATT or ATC, and their corresponding $h(\bm{x})$, weights, and target populations are in Supplementary Table \ref{tab:common_estimands}.
% One common causal estimand is the ATE, which averages $\tau(\bm{x})$ over the entire covariate population from which the sample was derived. For the ATE, $h(\bm{x}) = 1$ and the balancing weights are the IPW weights. Other common average casual estimands and their corresponding $h(\bm{x})$, weights, and target populations are in Web Table 1.
For a given $h(\bm{x})$, the sample estimator for $\tau_h$ is
 %\vspace{-0.2in}
%\begin{align*}
    $\hat{\tau}_h = \dfrac{\sum_i w_1(\bm{X}_i)Z_iY_i}{\sum_i w_1(\bm{X}_i)Z_i} - \dfrac{\sum_i w_0(\bm{X}_i)(1-Z_i)Y_i}{\sum_i w_0(\bm{X}_i)(1-Z_i)}$,
    %\label{eq:tau_estimator}
%\end{align*}
where $\hat{\tau}_h$ is consistent for $\tau_h$ when the weights in Equation \eqref{eq:w1} are used \citep{li2018}. For the following sections, aside from Section \ref{sec:var_prop}, we assume $\hat{\tau}_h$ is computed with estimated propensity scores as the true propensity scores are rarely known.

% \begin{table}[h!]
% \caption{ Common causal estimands and their corresponding $h(\bm{x})$, balancing weights 
% \citep{li2018}, and population.}
% \centering
% \begin{tabular}{cccc} \hline
% Estimand &  $h(\bm{x})$    &  $w_0(\bm{x}), w_1(\bm{x})$     &  Target Population \\ \hline
% ATE          & $1$     & $1/\{1-e(\bm{x})\}, 1/e(\bm{x})$       & Combined      \\
% ATT           & $e(\bm{x})$  & $e(\bm{x})/\{1-e(\bm{x})\}, 1$      & Treated      \\
% ATC         & $1-e(\bm{x}) $   & $1, e(\bm{x})/\{1-e(\bm{x})\}$   & Control      \\
% ATO          & $e(\bm{x})\{1-e(\bm{x})\}$& $e(\bm{x}), 1-e(\bm{x})$ & Overlap      \\   \hline    
% \end{tabular}
% \label{tab:common_estimands}
% \end{table}

\subsection{Decomposing sources of error in the presence of positivity violations}

In practice, one may need to mitigate variability and/or bias due to lack of overlap by modifying the estimand from the original estimand. We call the modified estimand a ``performance'' estimand, as it is often chosen pragmatically. In this section, we aim to characterize the sources of discrepancy between the original target estimand of interest and the specific estimator used in practice, which may or may not target a performance estimand. This characterization allows us to systematically navigate the bias and variance tradeoffs that account for both statistical errors and mismatch between the original estimand and performance estimand. 

%We propose two classes of estimands that may be incorporated within a causal analysis: population target estimands and performance estimands. 
We refer to the original estimand, which would typically correspond to a meaningful population and be viewed as the key estimand of scientific interest, as the population target estimand. %population target estimands are estimands such that $h(\bm{x})$ induces a meaningful population of interest and would typically be viewed as the key estimand of scientific interest. 
Performance estimands, on the other hand, are often chosen to provide improved statistical benefits (e.g. lower estimator bias, variance) in comparison to the population target estimand. The population target and performance estimands can be the same; for example, the ATO provides variance reduction compared to other estimands and may also correspond to a population of interest. However, we will generally consider estimands such as the ATO and those induced by propensity score trimming as performance estimands as these estimands were originally proposed to achieve estimators with reduced variance. Throughout the rest of the paper, we will consider the ATE ($h(\bm{x}) = 1$), $\tau$, as the population target estimand for demonstrative purposes, though the proposed methods and framing can straightforwardly apply to any target estimand.

Let $h_p(\bm{x})$ correspond to a performance estimand $\tau_{h_p}$ for a given analysis where the target estimand is $\tau$. Then the bias of $\hat{\tau}_{h_p}$ in relation to $\tau$ can be decomposed as
%\vspace{-0.2in}
\begin{align}
    E(\hat{\tau}_{h_p}) - \tau \:\: = \:\: \underbrace{E(\hat{\tau}_{h_p}) - \tau_{h_p}}_{\text{Statistical bias}}   \: \:\: +  \!\!\!\!
    \underbrace{\tau_{h_p} - \tau}_{\text{Estimand mismatch}}.
    \label{eq:bias_decomp}
\end{align}
% \begin{align}
%     E(\hat{\tau}_{h_p}) - \tau \:\: =\!\!\!\!\!\!\!\!\!\! \underbrace{\tau_{h_p} - \tau}_{(1) \text{ Estimand mismatch}} \!\!\!\!\!\! + \: \: \: \underbrace{E(\hat{\tau}_{h_p}) - \tau_{h_p}}_{(2) \text{ Statistical bias}}.
%     \label{eq:bias_decomp}
% \end{align}
The first term represents the statistical bias of the estimator $\hat{\tau}_{h_p}$, while the second term represents the mismatch between the ATE and the performance estimand. To minimize the discrepancy between $\hat{\tau}_{h_p}$ and the target estimand, both the estimand mismatch and the statistical bias of $\hat{\tau}_{h_p}$ must be controlled. Estimand mismatch and statistical bias are often at odds, especially in scenarios with substantial positivity violation; for example, $\tau_{h_p} = \tau$ would yield zero estimand mismatch, but $\hat{\tau}$ tends to have high statistical bias when there is low overlap. 
To reduce overall bias, it is desirable to identify a $\tau_{h_p}$ such that $\hat{\tau}_{h_p}$ has smaller statistical bias than $\hat{\tau}$ as well as limited estimand mismatch. 

\begin{figure}[h!]
    \centering
    \hspace*{-0.1in}
    \includegraphics[draft = false, scale = 0.5]{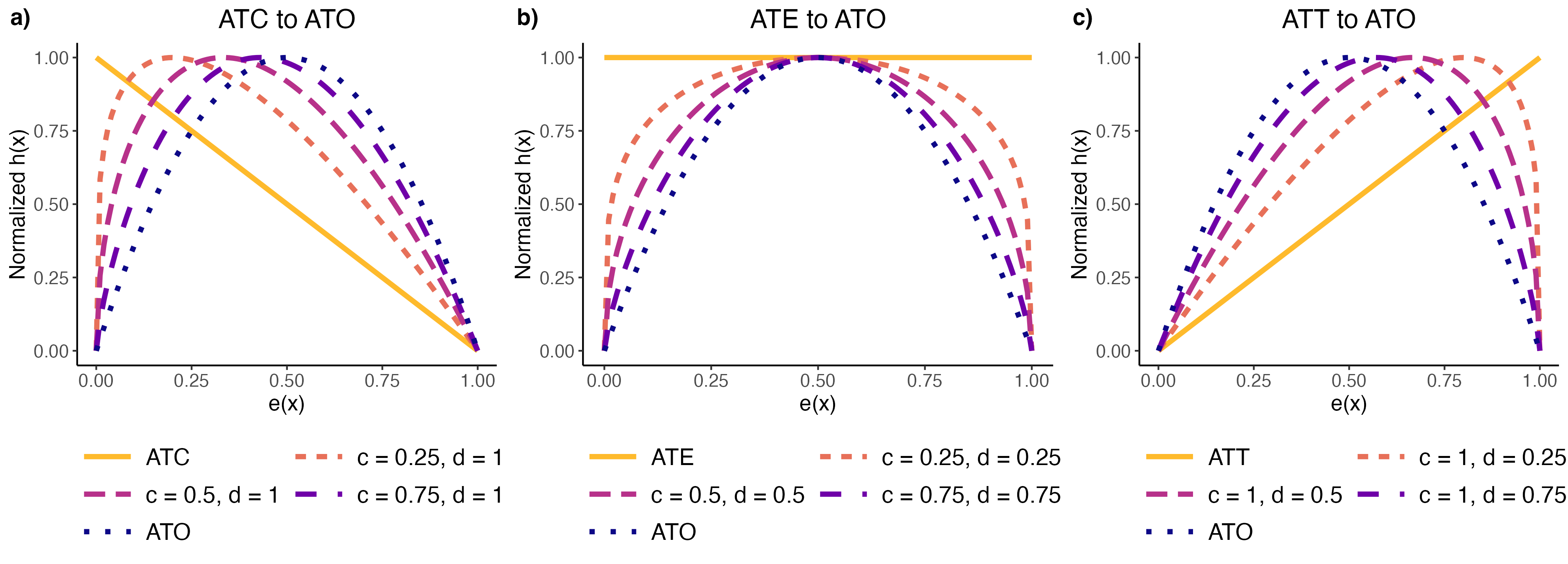}
    \caption{We show the normalized $h(\bm{x})$ in relation to the propensity score, $e(\bm{x})$, for a set of estimands. For each subplot, we plot $h(\bm{x})$ for a well-known estimand and the ATO and demonstrate how estimands within our proposed set (Equation \eqref{eq:pot_estimands}) are intermediaries between them.}
    \label{fig:potential_estimands}
\end{figure}

To help navigate the tradeoffs between estimand mismatch, statistical bias, and variance we propose an approach for the selection of the performance estimand, $\tau_{h_p}$. We allow for selection of $\tau_{h_p}$ from the following family of potential weight functions,
%\vspace{-0.2in}
\begin{equation}
    h_{c,d} (\bm{x})= e(\bm{x})^c\{1-e(\bm{x})\}^d \: ; \: 0 \leq c,d \leq 1,
    \label{eq:pot_estimands}
    %\vspace{-0.1in}
\end{equation}
where the ATE, ATC, ATT, and ATO are all special cases of this set (Figure \ref{fig:potential_estimands}). By varying the two parameters, $c$ and $d$, this family provides a smooth transition between each of these widely-used estimands. We can then select %an optimal 
$\tau_{h_p}$ from this set by characterizing the bias and variance of their corresponding estimators.

\subsection{Identifying measurable attributes of estimand mismatch and statistical bias}
To leverage the bias decomposition proposed in the previous section for estimand selection, it is necessary to understand the measurable attributes that impact estimand mismatch and statistical bias. We focus on the ATE as the target estimand, however, this work can also be applied to other estimands of interest such as the ATT, ATC, or when transporting the effect to a new population. 
For clarity in notation, we assume that the weights have been normalized such that $\sum_{i=1}^n w_{z}(\bm{X}_i)I(Z_i = z) = n_z$, the sample size of treatment group $z$. We can then express the difference between a performance estimator and the ATE as
\begin{align}
    \hat{\tau}_{h_p} - \tau = {} & \underbrace{\int \mu_1(\bm{x})d\{F_{n_0,0,\bm{w}_0} - F_n\}(\bm{x}) - \int \mu_0(\bm{x})d\{F_{n_1,1,\bm{w}_1} - F_n\}(\bm{x})}_{\text{Estimand mismatch}} \label{eq:estimand_mis}  \\
    &+\underbrace{\int \{\mu_1(\bm{x}) + \mu_0(\bm{x})\}d\{F_{n_1,1,\bm{w}_1} - F_{n_0,0,\bm{w}_0}\}(\bm{x})}_{\text{Statistical bias}} \label{eq:stat_bias}  \\
    &- \int \{\mu_1(\bm{x}) - \mu_0(\bm{x})\}d\{F - F_n\}(\bm{x}) \label{eq:zero_bias1} \\
    &+ \frac{1}{n_1}\sum_{i=1}^n w_{1}(\bm{X}_i)\epsilon_iZ_i - \frac{1}{n_0}\sum_{i=1}^n w_{0}(\bm{X}_i)\epsilon_i(1 - Z_i),
    \label{eq:zero_bias2}
\end{align}
where $\epsilon_i = Y_i(Z_i) - \mu_{Z_i}(\bm{X}_i)$, $F_n(\bm{x}) = \sum_{i=1}^n I(\bm{X}_i \leq \bm{x})/n$ is the empirical CDF, and $F_{n_z,z,\bm{w}_z}(\bm{x}) =\sum_{i=1}^n w_{z}(\bm{X}_i)I(\bm{X}_i \leq \bm{x}, Z_i = z)/n_z$ is the weighted empirical CDF. In this expression, term \eqref{eq:zero_bias1} goes to zero for a representative sample of the population and term \eqref{eq:zero_bias2} has mean zero. Then, the difference between our performance estimator and population target estimand can be characterized by terms \eqref{eq:estimand_mis} and \eqref{eq:stat_bias} which are induced by estimand mismatch and statistical bias, respectively. We note that term \eqref{eq:estimand_mis} often has non-zero expectation even with no estimand mismatch (the expectation is zero when using the true propensity scores); however, this term does tend to increase in size as estimand mismatch increases. The measurable quantities in terms \eqref{eq:estimand_mis} and \eqref{eq:stat_bias} are 1) the imbalance between the empirical CDFs $F_n$ and $F_{n_z,z,\bm{w}_z}$; and 2) the imbalance between empirical CDFs $F_{n,0,\bm{w}_0}$ and $F_{n,1,\bm{w}_1}$. To formally characterize estimand mismatch and statistical bias for a given $\hat{\tau}_{h_p}$, 
%one can then quantify the distance between empirical CDFs $F_n$ and $F_{n_z,z,\bm{w}_z}$ and empirical CDFs $F_{n_0,0,\bm{w}_0}$ and $F_{n_1,1,\bm{w}_1}$.  
we propose design-based metrics that measure these two distributional imbalances in Sections \ref{sec:mismatch_metric} and \ref{sec:stat_bias_metric}. To select %an optimal 
a performance estimand from the family of  $h(\bm{x})$ in Equation \eqref{eq:pot_estimands}, we evaluate $h_{c,d}(\bm{x})$ by these two metrics.

\subsection{Identifying estimands with minimal estimator variance}
\label{sec:var_prop}
While we have focused on estimator bias, we also often want to identify a performance estimand whose corresponding estimator has small or minimal variance. \cite{li2018} prove that overlap weights ($h(\bm{x}) = e(\bm{x})\{1-e(\bm{x})\}$) yield the estimator with the minimum asymptotic variance for all $\hat{\tau}_{h}$ under homoscedasticity. However, overlap weights do not necessarily yield the estimator with minimum asymptotic variance when homoscedasticity does not hold.
The following proposition establishes the variance conditions under which $\hat{\tau}_{h*}$ has minimum asymptotic variance (proof in Appendix A).  

\begin{proposition}
\it Let $v_z(\bm{x}) = Var\{Y(z)|\bm{X} = \bm{x}\}$ and $h^*(\bm{x}) > 0$ for all $\bm{x}$. Then if
%\vspace{-0.15in}
\begin{align*}
    % v_0(\bm{x}) &= v\frac{k_0(\bm{x})}{e(\bm{x})^{c}(1-e(\bm{x}))^{d-1}[k_0(\bm{x}) + k_1(\bm{x})]} \text{ and} \\
    % v_1(\bm{x}) &= v\frac{k_1(\bm{x})}{e(\bm{x})^{c-1}(1-e(\bm{x}))^{d}[k_0(\bm{x}) + k_1(\bm{x})]},
    v_0(\bm{x}) &= v\frac{k_0(\bm{x})\{1-e(\bm{x})\}}{h^*(\bm{x})\{k_0(\bm{x}) + k_1(\bm{x})\}} \text{ and}\:\:\: v_1(\bm{x}) = v\frac{k_1(\bm{x})e(\bm{x})}{h^*(\bm{x})\{k_0(\bm{x}) + k_1(\bm{x})\}},
\end{align*} $h(\bm{x}) \propto h^*(\bm{x})$ gives the minimum asymptotic variance of all balancing weight estimators $\hat{\tau}_h$ for $v \in \mathbb{R}^+$ and functions $k_0(\bm{x}), k_1(\bm{x}) > 0$ for all $\bm{x}$.  
\end{proposition}
% \textbf{Proposition 1} \textit{Let $v_z(\bm{x}) = Var\{Y(z)|\bm{X} = \bm{x}\}$ and $h^*(\bm{x}) > 0$ for all $\bm{x}$. Then if
% %\vspace{-0.15in}
% \begin{align*}
%     % v_0(\bm{x}) &= v\frac{k_0(\bm{x})}{e(\bm{x})^{c}(1-e(\bm{x}))^{d-1}[k_0(\bm{x}) + k_1(\bm{x})]} \text{ and} \\
%     % v_1(\bm{x}) &= v\frac{k_1(\bm{x})}{e(\bm{x})^{c-1}(1-e(\bm{x}))^{d}[k_0(\bm{x}) + k_1(\bm{x})]},
%     v_0(\bm{x}) &= v\frac{k_0(\bm{x})\{1-e(\bm{x})\}}{h^*(\bm{x})\{k_0(\bm{x}) + k_1(\bm{x})\}} \text{ and} \\
%     v_1(\bm{x}) &= v\frac{k_1(\bm{x})e(\bm{x})}{h^*(\bm{x})\{k_0(\bm{x}) + k_1(\bm{x})\}},
% \end{align*} $h(\bm{x}) \propto h^*(\bm{x})$ gives the minimum asymptotic variance of all balancing weight estimators $\hat{\tau}_h$ for $v \in \mathbb{R}^+$ and functions $k_0(\bm{x}), k_1(\bm{x}) > 0$ for all $\bm{x}$.} \\
This result further motivates selecting a performance estimand specific to a given dataset, as the overlap weights estimator may not have the minimum asymptotic variance let alone the minimum finite sample variance for some data. To characterize estimator variance and select %an optimal 
a performance estimand, we estimate the standard error of each $\hat{\tau}_{h_{c,d}}$ using a bootstrapped standard error estimator \citep{crump_dealing_2009}.

\section{Implementation of estimand selection framework}
\label{sec:methods}
\subsection{Characterizing estimand mismatch}
\label{sec:mismatch_metric}
We use the weighted energy distance as our measure of imbalance between distributions \citep{huling_energy_2024}. We choose the energy distance due to its performance in high dimensions and computational efficiency, however other measures of distributional imbalance such as the maximum mean discrepancy, Wasserstein distance, etc. could work well and be used here instead. The weighted energy distance between empirical CDFs $F_n$ and $F_{n_z,z,\bm{w}_z}$ is 
%\vspace{-0.2in}
\begin{align*}
    \mathcal{E}(F_{n_z,z,\bm{w}_z}, F_n) &= \frac{2}{n_zn}\sum_{i=1}^n\sum_{j=1}^n w_{z}(\bm{X}_i)I(Z_i = z)||\bm{X}_i -\bm{X}_j||_2 - \frac{1}{n^2}\sum_{i=1}^n\sum_{j=1}^n ||\bm{X}_i -\bm{X}_j||_2 \\
    &-\frac{1}{n_z^2}\sum_{i=1}^n\sum_{j=1}^n w_{z}(\bm{X}_i) w_{z}(\bm{X}_j)I(Z_i = Z_j = z)||\bm{X}_i - \bm{X}_j||_2. 
\end{align*}
In this context, $\mathcal{E}(F_{n_z,z,\bm{w}_z}, F_n)$ quantifies the difference between the population that is generated by $h_{c,d}(\bm{x})$ and the target ATE population. While non-zero energy
distances indicate the degree of difference between two distributions, the energy distance is
not a standardized metric and therefore is not comparable across different weight functions. To create a metric comparable across all $h_{c,d}(\bm{x})$, we construct a permutation-based statistical test for $H_0: F = F_{z,\bm{w}_z}$. The proposed test procedure for a given $F_n$ and $F_{n_z,z,w_z}$ follows:
%\vspace{-.08in}
\begin{enumerate}
    \item Compute the test statistic $\mathcal{E}(F_{n_z,z,\bm{w}_z}, F_n)$
    \item For $b = 1, \ldots, B$: i) Sample $F^{(b)}_{n_z}$ from $F_n$ without replacement; ii) Compute $\mathcal{E}(F^{(b)}_{n_z,\bm{w}_z}, F_n)$
    % \begin{enumerate}
    %     \item Sample $F^{(b)}_{n_z}$, of size $n_z$ from $F_n$ without replacement
    %     \item Compute $\mathcal{E}(F^{(b)}_{n_z,\bm{w}_z}, F_n)$
    % \end{enumerate}
    \item Compute the $p$-value, $p = \frac1{B}\sum_{b=1}^B I[\mathcal{E}(F_{n_z,z,\bm{w}_z}, F_n) \leq \mathcal{E}(F^{(b)}_{n_z,\bm{w}_z}, F_n)]$
\end{enumerate}
%Note that by construction, $\{F^{(b)}_{n_z}\}_{b=1}^B$ in step 2 approximates $F_z$ under $F = F_{z}$. Then, since $w_z$ has no relationship to the sampled $\{F^{(b)}_{n_z}\}_{b=1}^B$, $\{F^{(b)}_{n_z, \bm{w}_z}\}_{b=1}^B$ approximates $F_{z,\bm{w}_z}$ under the null $F = F_{z, \bm{w}_z}$; thus, $\{\mathcal{E}(F^{(b)}_{n_z,\bm{w}_z}, F_n)\}_{b=1}^B$ in step 2 approximates the test statistic under the desired null distribution. 
In step 2, since $\bm{w}_z$ has no relationship to the randomly sampled $\{F^{(b)}_{n_z}\}_{b=1}^B$, $\{F^{(b)}_{n_z, \bm{w}_z}\}_{b=1}^B$ approximates $F_{z,\bm{w}_z}$ under the desired null $F = F_{z, \bm{w}_z}$. We can use this $p$-value for inference or simply as a standardized measure of the imbalance between $F_{n_z,z,\bm{w}_z}$ and $F_n$, where smaller $p$-values tend to correspond to $\hat{\tau}_{h_{c,d}}$ with higher potential for large estimand mismatch. If we want to determine whether the population induced by $h_{c,d}(\bm{x})$ is different from the ATE population, we reject the null hypothesis that these populations are the same at $p < \alpha = 0.05$. 
%In practice, we calculate two $p$-values, one comparing each weighted treatment group $z$ to the targeted population. 
In Sections \ref{sec:simulation} and \ref{sec:real_data}, we use the minimum of the $p$-values computed for each treatment group $z$ as the main design-based metric for characterizing estimand mismatch.

\subsection{Characterizing statistical bias}
\label{sec:stat_bias_metric}
Let $G$ indicate the CDF for the fitted propensity scores $\hat{e}(\bm{X_i})$. The weighted energy distance between empirical CDFs $G_{n_0, 0, \bm{w}_0}$ and $G_{n_1, 1, \bm{w}_1}$ is
%\vspace{-0.15in}
{
\begin{align*}
   \mathcal{E}(G_{n_0,0,\bm{w}_0}, G_{n_1,1,\bm{w}_1}) &= \frac{2}{n_0n_1}\sum_{i=1}^n\sum_{j=1}^n w_{0}(\bm{X}_i)w_{1}(\bm{X}_j)I(Z_i = 0)I(Z_j = 1)||\hat{e}(\bm{X}_i) -\hat{e}(\bm{X}_j)||_2  \\
    &- \sum_{z=0}^1 \frac{1}{n_z^2}\sum_{i=1}^n\sum_{j=1}^n w_{z}(\bm{X}_i)w_{z}(\bm{X}_j)I(Z_i = Z_j = z)||\hat{e}(\bm{X}_i) - \hat{e}(\bm{X}_j)||_2. %\vspace{-0.1in}
    %&- \frac{1}{n_0^2}\sum_{i=1}^n\sum_{j=1}^n w_{0}(\hat{e}(\bm{X}_i))w_{0}(\hat{e}(\bm{X}_j))I(Z_i = Z_j = 0)||\hat{e}(\bm{X}_i) - \hat{e}(\bm{X}_j)||_2\\
    %&- \frac{1}{n_1^2}\sum_{i=1}^n\sum_{j=1}^n w_{1}(\hat{e}(\bm{X}_i)) w_{1}(\hat{e}(\bm{X}_j))I(Z_i = Z_j = 1)||\hat{e}(\bm{X}_i) -\hat{e}(\bm{X}_j)||_2.
\end{align*}}%
We use the propensity score empirical CDFs instead of the covariate empirical CDFs (as seen in term \eqref{eq:stat_bias})
because the use of propensities scores has been shown to be sufficient for covariate sample
balance \citep{rosenbaum1983} and to simplify computation. The permutation-based statistical test procedure for $H_0: G_{0, \bm{w}_0} = G_{1, \bm{w}_1}$ follows:
\begin{enumerate}
    \item Compute $\mathcal{E}(G_{n_0,0,\bm{w}_0}, G_{n_1,1,\bm{w}_1})$
    \item For $b = 1, \ldots, B$: i) Permute the treatment assignments to generate $G^{(b)}_{n_0}$ and $G^{(b)}_{n_1}$; ii) Compute $\mathcal{E}(G^{(b)}_{n_0}, G^{(b)}_{n_1})$
    % \begin{enumerate}
    %     \item Permute the treatment assignments to generate $G^{(b)}_{n_0}$ and $G^{(b)}_{n_1}$
    %     \item Compute $\mathcal{E}(G^{(b)}_{n_0}, G^{(b)}_{n_1})$
    % \end{enumerate}
    \item $p = \frac1{B}\sum_{b=1}^B I[\mathcal{E}(G_{n_0,0,\bm{w}_0}, G_{n_1,1,\bm{w}_1}) \leq \mathcal{E}(G^{(b)}_{n_0}, G^{(b)}_{n_1})]$
\end{enumerate}
%By permuting the treatment assignments in step 2, we break the relationship between treatment assignment and covariates such that $\{G_{n_0}^{(b)}, G_{n_1}^{(b)}\}_{b=1}^B$ approximate $G_0, G_1$ under the null distribution $G_0 = G_1$ which is approximately the $G_{0, \bm{w}_0} = G_{1, \bm{w}_1}$ null distribution through treatment assignment permutation. Then, $\{\mathcal{E}(G^{(b)}_{n_0}, G^{(b)}_{n_1})\}_{b=1}^B$ approximates the test statistic under $H_0$. 
In step 2, permuting the treatment assignment breaks the relationship between treatment assignment and covariates such that $\{G_{n_0}^{(b)}, G_{n_1}^{(b)}\}_{b=1}^B$ approximate $G_0, G_1$ under $G_0 = G_1$ which is equivalent to the desired null $G_{0, \bm{w}_0} = G_{1, \bm{w}_1}$ in population. Here, we primarily use $p$ as a standardized measure of the imbalance between $G_{n_0,0,\bm{w}_0}$ and $ G_{n_1,1,\bm{w}_1}$, where smaller $p$-values tend to correspond to $\hat{\tau}_{h_{c,d}}$ with the potential for larger statistical bias.  We are less interested in specifying an $\alpha$ level as $\bm{w}_0$, $\bm{w}_1$ are designed to make these distributions equal and thus we expect these $p$-values to generally be high. While this $p$-value is motivated by characterizing statistical bias due to lack of overlap, in practice it characterizes statistical bias due to any source(s), such as propensity model misspecification.

\subsection{Estimand selection}
\label{sec:methods_estimand_selec}

%There are many potential ways to select an estimand for a dataset given the two proposed metrics for characterizing bias and the bootstrapped standard error estimator. 
We propose the following selection procedure as an approach for enumerating a sequence of estimators, each approximately optimal (i.e., low potential for statistical bias and low estimator variance) for a given subjective choice in emphasis of estimand mismatch versus statistical bias. Thus, the procedure allows the analyst to choose the point along this spectrum most appropriate for their research question. The procedure is as follows:
\begin{enumerate}
    \item Compute $p$ as defined in Sections \ref{sec:mismatch_metric} and \ref{sec:stat_bias_metric} and the bootstrapped standard error estimator for some grid of potential $h_{c,d}(\bm{x})$, $c,d \in [0,1]$. %(in Section \ref{sec:simulation} we use all combinations of $c,d = 0, 0.05, 0.10, \ldots, 1$).
    \item For each set of $p$-values, (i.e., corresponding to estimand mismatch and statistical bias) create density contours on the $(c,d) \in [0,1] \times [0,1]$ space (Supplementary Figure \ref{fig:example_contour}a) %\ref{fig:example_contour}a) %Interpolation may be needed to smooth the contours. 
    and use kriging to interpolate the estimated standard errors across a finer grid. %over $[0, 1] \times [0,1]$.
    \item Intersect the estimand mismatch and statistical bias $p$-value contours. For each estimand mismatch $p$-value contour level, identify the area intersecting with the largest statistical bias $p$-value contour level and select the $h_{c,d}(\bm{x})$ with the smallest estimated standard error in that area (Supplementary Figure \ref{fig:example_contour}b)).%\ref{fig:example_contour}b).
    %\item Use kriging to interpolate the estimated standard errors across a finer grid over $[0, 1] \times [0,1]$. For each estimand mismatch $p$-value area from the previous step, select the $h_{c,d}$ with the smallest estimated standard error. 
\end{enumerate}

The first step in the procedure characterizes a grid of estimands by the proposed estimand mismatch and statistical bias metrics as well as estimated standard error.
%; finer grids take longer to compute but provide more complete information. 
The second step simply extends this characterization across either all estimands defined by Equation \eqref{eq:pot_estimands} or a finer grid of estimands. Step three identifies and selects an estimand with low potential for statistical bias and the smallest estimated standard error within each estimand mismatch $p$-value contour. This results in a sequence of 
%optimal (i.e., low
%potential for statistical bias and low variance) 
estimands; as the estimand mismatch $p$-value contour level gets smaller, the selected estimand tends to have larger estimand mismatch but smaller estimator statistical bias and variance. 
%While the two $p$-value metrics are design-based, the standard error estimator uses outcome data; we discuss a potential way to modify the procedure such that it is entirely design-based in Section 6.

In general, we recommend selecting the %optimal 
estimand identified within the $(0.05, 0.1]$ $p$-value contour, as the corresponding estimator will tend to have the lowest standard error among estimands where we fail to reject the null hypothesis that the weighted estimand population is the same as the ATE population. Although we propose this procedure and implement it within Sections \ref{sec:simulation} and \ref{sec:real_data}, it can be varied depending on the particular data application. For example, if only the statistical performance of the estimator is of interest, one could select the estimand in the largest statistical bias $p$-value contour with the smallest estimated standard error.
Ultimately, the selection procedure can be adapted based on the perceived importance of estimand mismatch, statistical bias, and estimator variance in a given application. 
%\vspace{-0.15in}
\section{Simulation experiments}
\label{sec:simulation}
\subsection{Methods}
We construct data generating scenarios with varying levels of propensity score overlap, treatment effect heterogeneity, and percent of treated individuals. Through these scenarios, we aim to validate the metrics proposed in Sections \ref{sec:mismatch_metric} and \ref{sec:stat_bias_metric} across a variety of data generating processes. %We further aim to determine whether the ability of these metrics to characterize estimand mismatch and statistical bias varies across data generating scenarios. 
In addition, we aim to illustrate the scenarios in which our estimand selection procedure can identify estimands whose corresponding estimators have benefits over the ATE and ATO estimators. %By exploring different levels of propensity score overlap and treatment effect heterogeneity, 
We also demonstrate how the different data generation attributes influence the bias-variance tradeoffs of estimand selection within our procedure.

\subsubsection{Data generation and characterizing estimands}
\label{sec:char_sim_methods}
We implement nearly the same simulation data generation process as in \cite{li2019}. Here, we drop the observation index on variables for clarity of presentation. We generate $V_1, \ldots , V_6$ from a multivariate normal distribution with $E[V_i] = 0$, $Var(V_i) = 1$ for all $i = 1, \ldots, 6$ and $Cov(V_i, V_j) = 0.5$ for all $i \neq j$. We take $X_1\textrm{--}X_3 = V_1\textrm{--}V_3$ and dichotomize $V_4\textrm{--}V_6$ such that $X_j = I[V_j < 0]$ for $j = 4, 5,6$. The true propensity scores  are generated as $e(\bm{X}) = \{1 + \exp(-\alpha_0 - \sum_{j=1}^6 \alpha_j X_j)\}^{-1}.$ Our continuous outcome, $Y$, is generated as $Y \sim N\{E(Y|Z, \bm{X}), 1\}$ where $E(Y|Z, \bm{X}) = \beta_0 + \sum_{j=1}^6 \beta_j X_j + \Delta(\bm{X}) Z$. Similar to \cite{mao2019}, we choose two treatment effect functions that have a concave relationship with $e(\bm{X})$. The two functions have different ranges that represent the level of treatment effect heterogeneity, 1) Medium: $\Delta(\bm{x}) = 8e(\bm{x})\{1-e(\bm{x})\} $ and 2) High: $\Delta(\bm{x}) = 16e(\bm{x})\{1-e(\bm{x})\} -1$. 
%that have a concave relationship with the true propensity scores.
% \begin{align}
%     \Delta(\bm{x}) &= 8e(\bm{x})\{1-e(\bm{x})\} \label{eq:delta1}\\
%     \Delta(\bm{x}) &= 16e(\bm{x})\{1-e(\bm{x})\} -1. \label{eq:delta2}
% \end{align}
%Both functions are centered at one, but the second function has a larger range such that the functions represent medium and high treatment effect heterogeneity, respectively. 
%but the first function has range $(0,2)$ while the second has range $(-1,3)$; therefore, these functions represent a medium and high treatment effect heterogeneity, respectively. 
Consistent with \cite{li2019}, the parameters for the true propensity score model are $(\alpha_1, \alpha_2, \alpha_3, \alpha_4, \alpha_5, \alpha_6) = (0.15\gamma, 0.3\gamma, 0.3\gamma, -0.2\gamma, -0.25\gamma, -0.25\gamma)$ and $\alpha_0$ is chosen to generate the desired percent of treated individuals ($25\%$ and $50\%$). We explore $\gamma = 1, 2, 3, 4$ where $\gamma = 1$ generates data with substantial propensity score overlap and $\gamma = 4$ generates data with minimal propensity score overlap and substantial density of propensity scores in the tails (Supplementary Figure \ref{fig:ps_overlap}). 
%We choose $\alpha_0$ to generate the desired percent of treated individuals in the data ($25\%$ and $50\%$). 
 For our outcome model, we set $(\beta_0,\beta_1, \beta_2, \beta_3, \beta_4, \beta_5, \beta_6) = (0, -0.5, -0.5, -1.5, 0.8, 0.8, 1)$. 

 We have a total of 16 simulation scenarios that represent all combinations of $\gamma = 1,2,3,4$, $25\%$ and $50\%$ treated individuals, and medium and high treatment effect heterogeneity. For each simulation scenario we generate 1000 independent datasets with 1000 observations each. For each simulated dataset, we use all combinations of $c,d = 0, 0.05, 0.1, \ldots, 1$ as the grid of candidate estimand values ($441$ estimands). We calculate the true value of each estimand using 10 million Monte Carlo samples. For each estimand, we calculate the average treatment effect estimate, absolute percent statistical bias, %the average estimated standard error of the estimator, empirical standard deviation, 
 and the average $p$-values from the two proposed metrics (Sections $\ref{sec:mismatch_metric}$ and $\ref{sec:stat_bias_metric}$). 
 %For our measure of distributional imbalance between $F_n$ and $F_{n,z,w}$, we calculate two $p$-values, one comparing each weighted treatment group to the targeted population. For all simulation results, we use the minimum of these two $p$-values as the main design-based metric for characterizing estimand mismatch.

\subsubsection{Estimand selection}
\label{sec:selec_sim_methods}
Using the results from \ref{sec:char_sim_methods}, we select a sequence of %optimal 
estimands for each simulated dataset with our proposed estimand selection procedure (Section \ref{sec:methods_estimand_selec}). We use cubic spline interpolation to smooth the $p$-value contours and interpolate standard error estimates over a $500 \times 500$ grid using kriging with a Gaussian variogram model. With our procedure, the number of selected estimands varies by the range of estimand mismatch $p$-values in the data; thus, for each estimand mismatch $p$-value contour level, only a subset of the 1000 datasets have a corresponding 
selected estimand. Since the set of datasets that have a specific range of estimand mismatch $p$-values may differ systematically, we select two different sample size cutoffs ($n= 500,900$) to compute aggregate metrics over.
% Since our procedure selects an estimand for each estimand mismatch contour level, the number of selected estimands varies by the range of estimand mismatch $p$-values in the data. 
% Therefore, for each estimand mismatch $p$-value contour level, only a subset of the 1000 datasets have a corresponding %optimal
% selected estimand; the larger the $p$-value contour level, the fewer datasets that have %an optimal 
% an estimand selected within that contour. Since the set of datasets that have a specific range of estimand mismatch $p$-values may differ systematically, we select two different sample size cutoffs ($n= 500,900$) to compute aggregate metrics over. 
%Effectively, we 1) determine the largest $p$-value contour level, $p^*$, with at least $n$ datasets with %optimal 
%selected estimands; 2) identify the specific datasets with 
%optimal 
%selected estimands in the $p^*$ contour; and 3) compute aggregate metrics for all $p$-value contour levels with $p < p^*$ over the datasets identified in step 2.
We calculate the MSE with respect to the true ATE for the ATE estimator, the estimators corresponding to estimands selected with our method, and the ATO estimator.

\subsection{Results}
\subsubsection{Characterizing estimand mismatch}

\begin{figure}[h!]
    \centering
    \hspace*{-0.2in}
    \includegraphics[scale = 0.8]{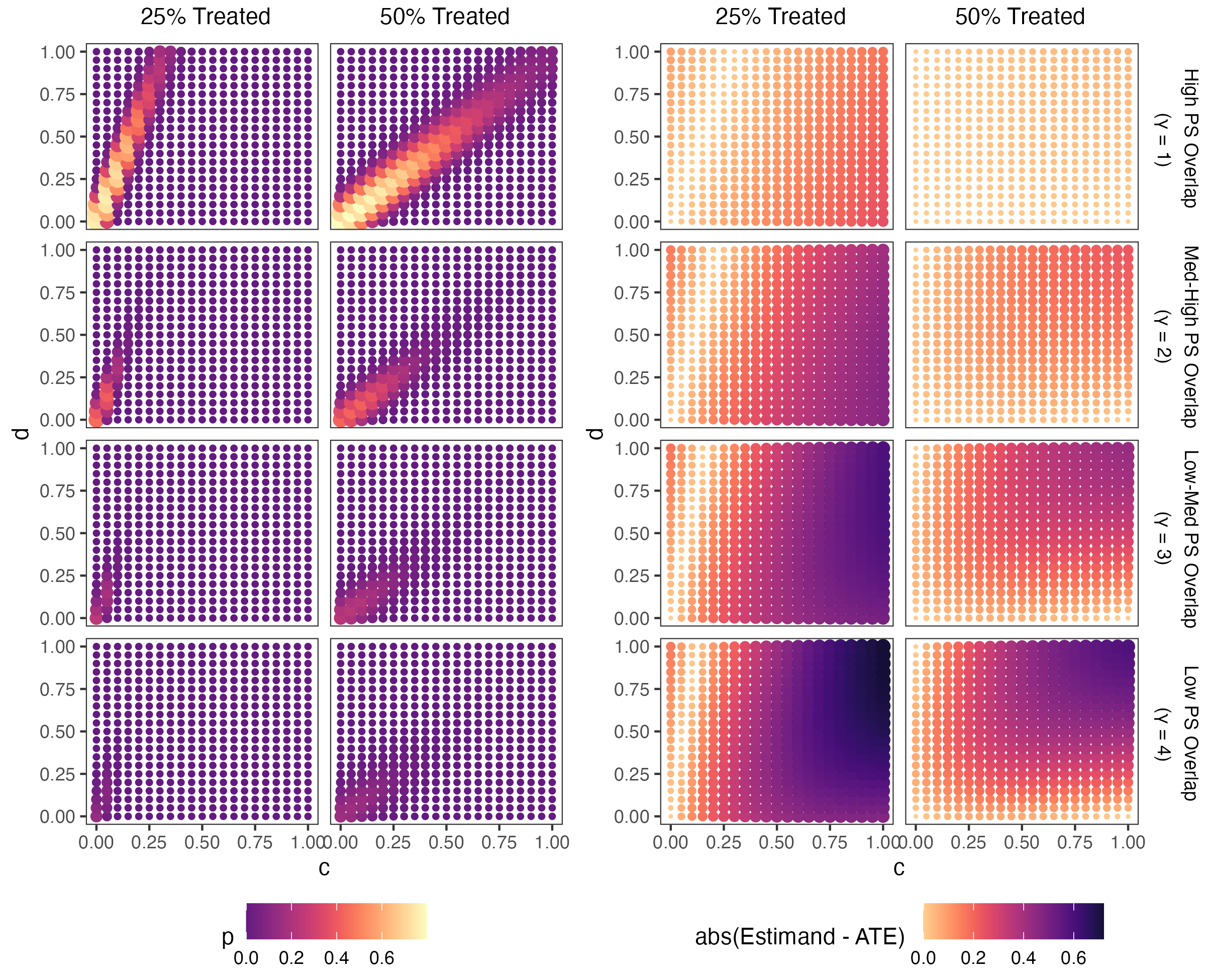}
    \caption{The left panel presents the average of the minimum of the two estimand mismatch related $p$-values, calculated as described in Section \ref{sec:mismatch_metric}, for each estimand. The right panel presents the absolute difference between the true estimands and the true ATE (i.e., the absolute estimand mismatch) under the medium $\Delta$ heterogeneity scenario. Columns distinguish the 25\% and 50\% treated scenarios while rows distinguish the different propensity score overlap scenarios.}
    \label{fig:p_cov}
\end{figure}

For the 50\% treated scenarios, the estimands with $p > 0.05$ (i.e., the estimands whose weighted covariate populations are similar to the ATE covariate population) follow the region along the diagonal ($c=d$) which, at high propensity score overlap, goes between the ATE to the ATO (Figure \ref{fig:p_cov}, left panel). For all of these estimands, $h_{c,d}(\bm{x})$ peaks around propensity scores of $0.5$. In contrast, for the 25\% treated scenarios, the estimands with $p > 0.05$ follow a trend above the diagonal ($c < d$) such that $h_{c,d}(\bm{x})$ peaks at propensity scores around $0.25$. Therefore, estimands that induce a covariate population similar to the ATE population are estimands where $h_{c,d}(\bm{x})$ peaks at the proportion of treated individuals. As propensity score overlap decreases, fewer estimands yield $p > 0.05$ and these estimands tend to have $c,d$ close to the ATE (e.g. $c, d < 0.25$). 

 For the 25\% treated scenarios, there is a clear relationship between the $p$-value metric and absolute estimand mismatch; estimand mismatch tends to be smallest along a trend above the diagonal (Figure \ref{fig:p_cov}, right panel). However, for the 50\% treated scenarios, this relationship is not as immediately clear. Estimand mismatch depends not only on distributional imbalance, but also on how the treatment effect varies across the covariate distribution (see term \eqref{eq:estimand_mis}). Therefore, $p < 0.05$ from our proposed metric indicates estimands with the \textit{potential} for large absolute estimand mismatch, but the exact relationship will vary by data generating mechanism. When data is generated with a treatment effect that is a linear function of $e(\bm{X})$, there is clearer visual relationship between the $p$-value metric and estimand mismatch for the 50\% treated scenarios; the estimands with the smallest estimand mismatch follow the diagonal (Supplementary Figure \ref{fig:p_cov_supp}).
 %Estimands close to the ATE do tend to have low estimand mismatch, however, many estimands further from the ATE along the diagonal have comparatively large estimand mismatch. Estimand mismatch depends not only on distributional imbalance, but also on how the treatment effect varies across the covariate distribution (see term \eqref{eq:estimand_mis}). Therefore, $p < 0.05$ from our proposed metric indicates estimands with the \textit{potential} for large absolute estimand mismatch, but the exact relationship between $h_{c,d}(\bm{x})$ and estimand mismatch will vary by data generating mechanism.
 %the specific estimands with large estimand mismatch vary by data generating mechanism. 
 %When we generate the data such that the treatment effect has a linear, rather than quadratic, relationship with the propensity score, the estimands with the smallest estimand mismatch follow the diagonal regardless of how far these estimands get from the ATE (Web Figure 3).%\ref{fig:p_cov_supp}).
 %Web Figure \ref{fig:p_cov_supp} presents absolute estimand mismatch when the treatment effect has a linear relationship with the propensity score, rather than quadratic. For this data generating process with 50\% treated individuals, the estimands with the smallest estimand mismatch follow the diagonal regardless of how far these estimands get from the ATE.
 
 For all 25\% treated data generating scenarios, 
 %shown in Figure \ref{fig:p_cov} and Web Figure 3, %\ref{fig:p_cov_supp}, 
 no estimands with absolute estimand mismatch above the 0.20 quantile %(within a data generating scenario)
 have $p > 0.05$. For 50\% treated scenarios, no estimands with an absolute estimand mismatch above the 0.45 quantile have $p > 0.05$, except for the $\gamma = 1,2$ scenarios from Figure \ref{fig:p_cov}. However, in these  scenarios the absolute estimand mismatch is relatively small and therefore selecting any of the proposed estimands will yield low estimand mismatch. Thus, the $p$-value metric protects against selecting estimands with high estimand mismatch across a variety of data generating mechanisms.

 \subsubsection{Characterizing statistical bias}
\label{sec:char_sim_res_statbias}
\begin{figure}[h!]
    \centering
    \hspace*{-0.2in}
    \includegraphics[scale = 0.8]{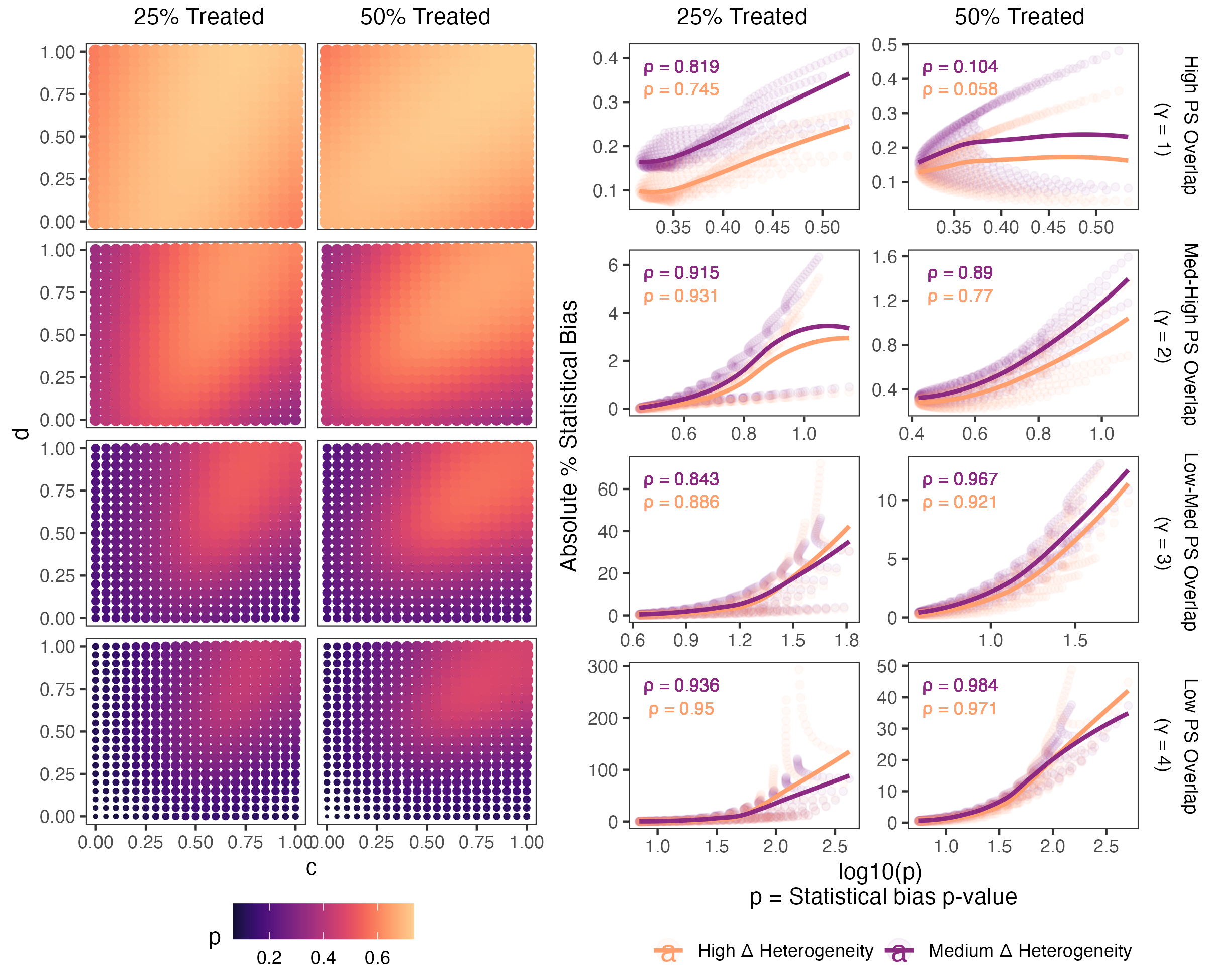}
    \caption{The left panel presents the average $p$-value corresponding to the statistical bias metric, calculated as described in Section \ref{sec:stat_bias_metric}, for each estimand. The right panel presents the relationship and Spearman correlation coefficient ($\rho$) between $\log_{10}(p)$ and the absolute percent statistical bias of the estimator for each estimand. Columns distinguish the 25\% and 50\% treated scenarios while rows distinguish the different propensity score overlap scenarios.}
    \label{fig:p_ps}
\end{figure}

For both proportion treated scenarios, the $p$-values characterizing statistical bias tend to be largest at $c,d > 0.75$, close to the ATO (Figure \ref{fig:p_ps}, left panel). As propensity score overlap decreases the $p$-values get smaller, indicating that the estimators have a higher potential for statistical bias. For the 50\% treated scenarios, the region of higher $p$-values tends to move from the ATO to the ATE along the diagonal.
%, with a majority of these estimands being closer to ATO. While most of the higher $p$-value estimands are also close to the ATO 
For the 25\% treated scenarios, estimands above the diagonal ($c < d$) tend to have lower $p$-values (and therefore a higher potential for statistical bias)  relative to the 50\% treated scenario.

There is a strong relationship between the $p$-values and absolute percent statistical bias, as measured by Spearman's correlation coefficient (Figure \ref{fig:p_ps}, right panel). The correlation tends to increase as propensity score overlap decreases; thus, as the potential for statistical bias gets higher, the $p$-value metric has a stronger relationship with statistical bias. Only in one scenario (50\% treated, $\gamma = 1$) is there a weak relationship between the $p$-value and absolute percent statistical bias, however, all the estimands in this scenario have small statistical bias ($< 0.5\%$ bias). We find that estimands with $p < 0.30$ are at risk of substantial bias, with at least $28.6\%$ of corresponding estimators having absolute percent statistical bias greater than $5\%$. While the exact numerical relationship between the $p$-value and absolute percent statistical bias will differ by data generating scenario, statistical bias tends to increases considerably as $p$ gets close to zero (Supplementary Figure \ref{fig:p_ps_supp}).
%We also want to identify at what $p$-values the estimators are generally at higher risk of substantial statistical bias. For estimands (across all data generating scenarios) with $p \in (0.25, .30]$, $28.6\%$ of their corresponding estimators have absolute percent statistical bias greater $5\%$ (Web Figure \ref{fig:p_ps_supp}). For estimands with $p \in (0.20, 0.25]$, $23.2\%$ of their corresponding estimators have absolute percent statistical bias greater than $10\%$. While the exact numerical relationship between the $p$-value and absolute percent bias will differ by data generating scenario, we find that estimands with $p < 0.30$ are at high risk of substantial statistical bias and that statistical bias tends to increases considerably as $p$ gets closer to zero. 

% \subsubsection{Standard error of estimators}
% We simulate the data to have homoscedastic residual variance such that our results are consistent with \cite{li2018}; the estimand corresponding to the minimum variance estimator tend to be close to the ATO ($c=d=1$) regardless of data generating scenario (Web Figure \ref{fig:se}). However, as shown in Proposition 1, this relationship could change when the residual variance is not homoscedastic. Of additional note, lack of propensity score overlap also leads to bias in variance estimation. When there is low propensity score overlap, the average standard error (SE) tends to underestimate the true SE of the estimators (as measured by the empirical standard deviation (SD)) (Web Figure \ref{fig:se_emp_sd}). \cite{mao2019} found the same relationship when using the sandwich variance estimator in this setting. Despite this discrepancy, there is a very strong relationship between the \textit{relative} ranking of the average estimated SE and empirical SD (all but one scenario has $\rho > 0.985$) across all data generating scenarios. This indicates that the bootstrapped SE estimator is still suitable for comparing and selecting estimands with minimal or small variance for a given dataset.

\subsubsection{Estimand selection}
\label{sec:selec_sim_res}

\begin{figure}[h!]
    \centering
    \hspace*{-0.3in}
\includegraphics[scale = 0.7]{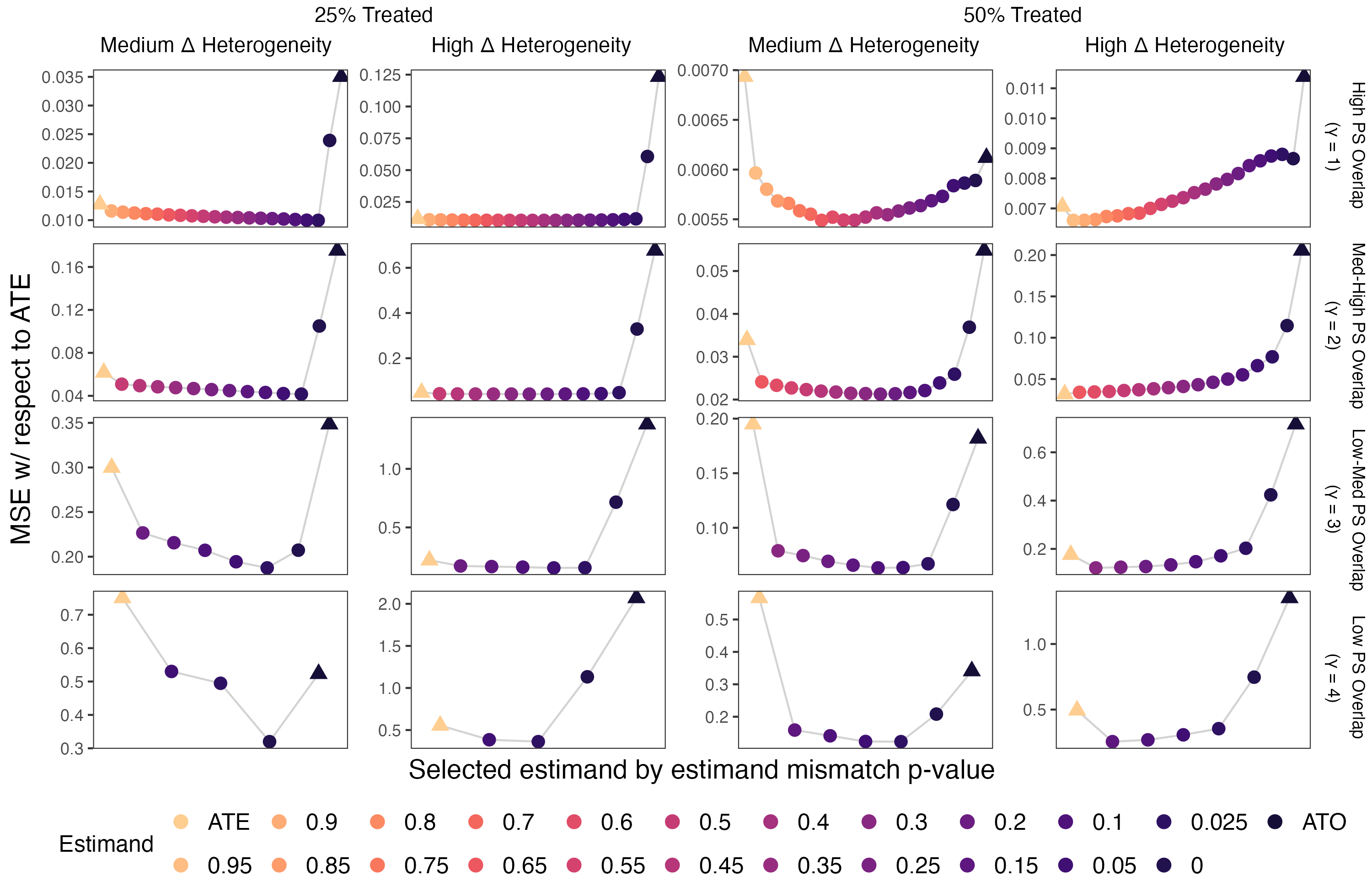}
    \caption{Mean squared error (MSE) with respect to the true ATE for the estimators for the ATE, the estimands selected with our procedure, and the ATO.  Each of the 16 subplots corresponds to a single simulation scenario as labeled by the top and right axes. Estimands selected from our framework are labeled by the lower bound of their corresponding estimand mismatch $p$-value contour. From left to right in each subplot, the depicted estimands become less similar the ATE and more similar to the ATO. Estimands selected (i.e., estimand mismatch $p$-value contours present) in at least 500 of the simulated datasets are shown.}
    \label{fig:mse_500n}
\end{figure}

Across data generating scenarios (for a $n = 500$ cutoff), the estimators corresponding to estimands selected with our method have a lower mean squared error (MSE) with respect to the ATE than either the ATE or ATO estimators (Figure \ref{fig:mse_500n}).
%; this indicates that our procedure can identify estimands that have better overall performance than simply targeting either extremal estimand. 
%From left to right in each subplot of Figure \ref{fig:mse_500n}, the points depict estimands that become less similar the ATE and more similar to the ATO. 
The curves in Figure \ref{fig:mse_500n}, specifically within low-medium propensity score overlap ($\gamma = 3$) scenarios, help demonstrate the bias-variance tradeoff when the scientific interest is with respect to the ATE. %rather than a performance estimand. 
For medium $\Delta$ heterogeneity, both the ATE (tends to have large variance) and ATO (tends to have substantial bias due to estimand mismatch) estimators have a large MSE indicating that bias and variance contribute relatively equally to the MSE. Thus, the estimators selected by our method all have lower MSE than the ATE and ATO estimators by moderating the size of both bias and variance. In contrast, for high $\Delta$ heterogeneity, the ATO estimator has a much larger MSE than the ATE estimator, indicating that bias (specifically estimand mismatch) dominates the MSE. Thus, estimators corresponding to larger estimand mismatch $p$-value contour levels tend to have the smallest MSE as they have reduced statistical bias and variance compared to the ATE estimator without incurring substantial estimand mismatch. We identify similar trends with a sample size cutoff of $n=900$ (Supplementary Figure \ref{fig:mse_900n}). 
  The average estimand per estimand mismatch $p$-value contour level, across all datasets, is shown in Supplementary Figure \ref{fig:avg_est}.
%to demonstrate on average which $c,d$ values are selected.
Consequently, there is a clear intuition for selecting a single estimand from the sequence of estimands identified through our method. %While a default estimand selection could be the estimand selected within the $(0.05, 0.1]$ $p$-value contour, one can use both subject knowledge of the expected treatment effect heterogeneity and preferences with respect to variance and bias to select a single estimand.
If treatment effect heterogeneity is expected to be low, the estimands from smaller $p$-value contours will tend to minimize the MSE; if treatment effect heterogeneity is expected to be high, the estimands from larger $p$-value contours will tend to minimize the MSE. For applications where variance reduction has higher priority than retaining the target estimand, either the ATO or the estimands from the smallest $p-$value contours will tend to be superior estimand choices. For applications where bias reduction with respect to the target estimand is the priority, estimands from larger $p$-value contours will tend to perform the best. In general, estimands selected from our method tend to have lower MSE than either the ATE or ATO because they are intermediate options within the bias-variance tradeoff; they tend to mitigate the size of statistical bias, estimand mismatch, and variance rather than minimizing just one of these attributes.

\section{Application to right heart catheterization data}
\label{sec:real_data}
Right heart catheterization (RHC) is an invasive medical procedure that is performed on critically ill patients in order to measure cardiac function. An observational study by \cite{connors1996} examined the impact of RHC on survival for patients in the ICU. The study was conducted between 1989-1994 across five medical centers within the United States. To account for treatment selection bias within the data, the authors matched treated and untreated individuals by propensity score matching as part of their analysis (\cite{rosenbaum_assessing_1983b, rosenbaum1983}). The data has also been re-analyzed in various other works \citep{hirano2001, crump_dealing_2009, rosenbaum_optimal_2012, li2018, huling_energy_2024}. The primary analysis showed that RHC may be detrimental to survival, which was contrary to the popular belief that RHC improved patient outcomes. 

The data consists of 5735 individuals, with 2184 treated (RHC applied within 24 hours of hospital admission) and 3551 controls. The outcome of interest is an indicator of survival at 30 days and there are 71 covariates selected by a panel of experts as variables that relate to practitioners' decision about whether to use RHC (\cite{connors1996}). We estimate the propensity score using a logistic regression with all covariates and age squared, as both younger and older patients are less likely to receive RHC. There is low to moderate overlap in the estimated propensity scores of the treated and control groups and many propensity scores are close to zero (Supplementary Figure \ref{fig:rhc_ps_dist}). 
This indicates that there is potential for poor statistical performance of the ATE estimator. When applying IPW weights, the average absolute standardized mean difference (SMD) of treated and control covariates after weighting is $0.018$ and the maximum is $0.062$; in general, the treated and control groups are relatively balanced after weighting (Supplementary Figure \ref{fig:rhc_cov_balance}).
%shows the 20 most imbalanced covariates after weighting, however in general the treated and control groups are relatively balanced after weighting.
For illustrative purposes, we quantitatively assess treatment effect heterogeneity through a series of linear probability models and find minimal evidence that there is substantial treatment effect heterogeneity within the data. In practice, this assessment would impact downstream inference. We apply our methods as described in Section \ref{sec:methods} to characterize and select a sequence of estimands. We use a grid of all combinations $c,d = 0, 0.05, 0.10, \ldots, 1$ and $1,000$ null distribution or bootstrap replications for all metrics.

\begin{figure}[tb]
    \centering
    \includegraphics[scale = 0.85]{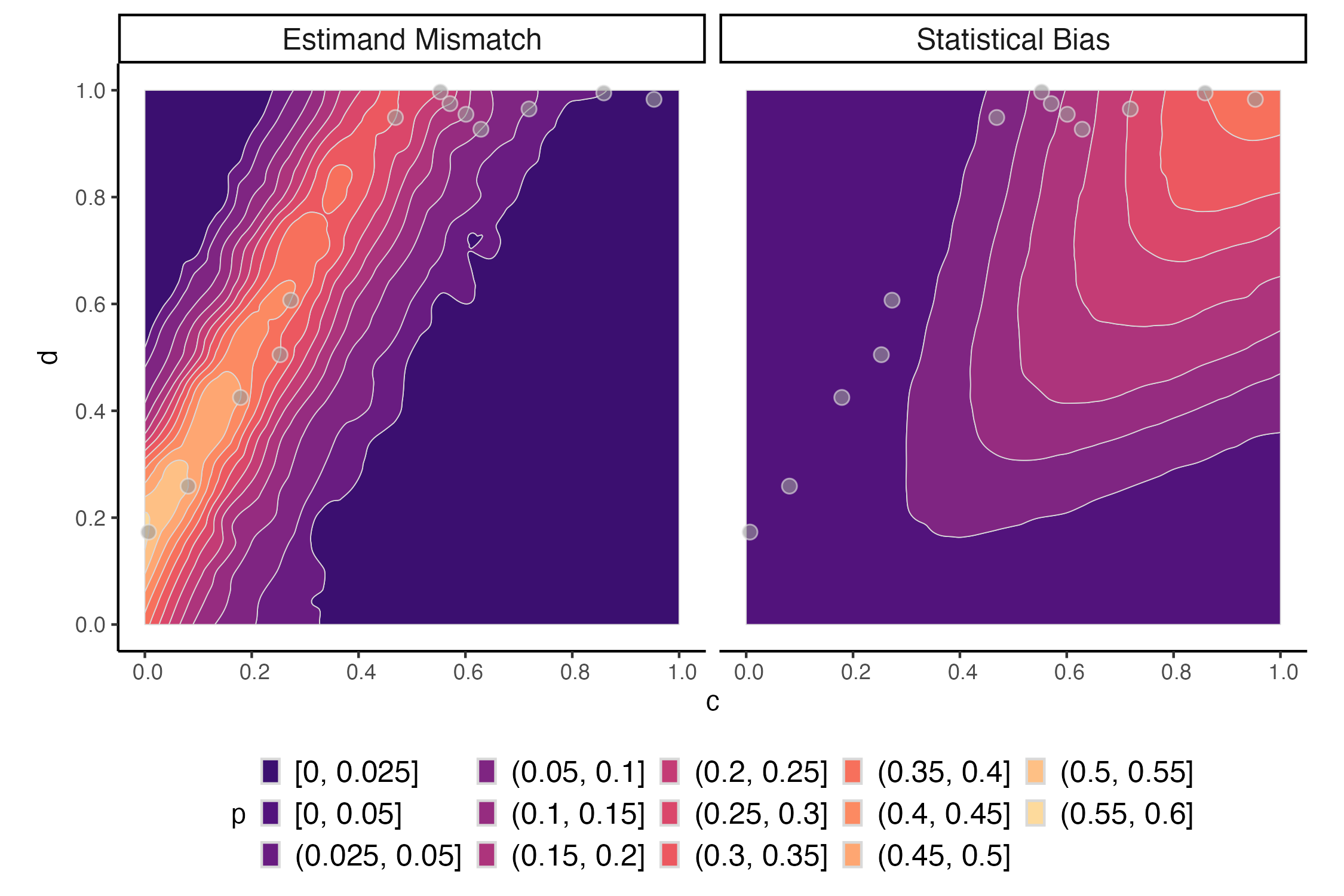}
    \caption{Contours of the $p$-values corresponding to estimand mismatch and statistical bias for the right heart catheterization study. Points on the figures show the %optimal
    estimands selected by our method (Section \ref{sec:methods_estimand_selec}) for each estimand mismatch $p$-value contour. Note that the estimand mismatch panel has contours of $[0, 0.025]$ and $(0.025, 0.05]$ while the statistical bias panel only has the contour $[0, 0.05]$.}
    \label{fig:contours_rhc}
\end{figure}

%Figure \ref{fig:contours_rhc} shows the estimand mismatch and statistical bias $p$-value contours for this data. 
For the $p$-values pertaining to estimand mismatch, the estimands with $p > 0.05$ generally follow a linear trend above the diagonal on the plot; for these estimands, $h(\bm{x})$ peaks around propensity scores of $0.38$, the proportion of treated individuals in the data (Figure \ref{fig:contours_rhc}). The $p$-values for characterizing statistical bias are largest for the estimands in the top right corner of the plot around the ATO. Of note, the statistical bias $p$-values are quite small (between $0$ and $0.40$); only only estimands with $c,d \geq 0.8$ have $p$-values greater than $0.30$. This indicates that the estimators for many of the estimands have a high potential for statistical bias. 
%- especially for the large region of estimands that fall within the $[0, 0.05]$ $p$-value contour. 
In comparison to simulations, here we do not know the true propensity score model, and therefore the $p$-values characterize the potential for statistical bias both from the lack of overlap and any propensity model misspecification. 

\begin{figure}[tb]
    \centering
    \includegraphics[scale = 0.8]{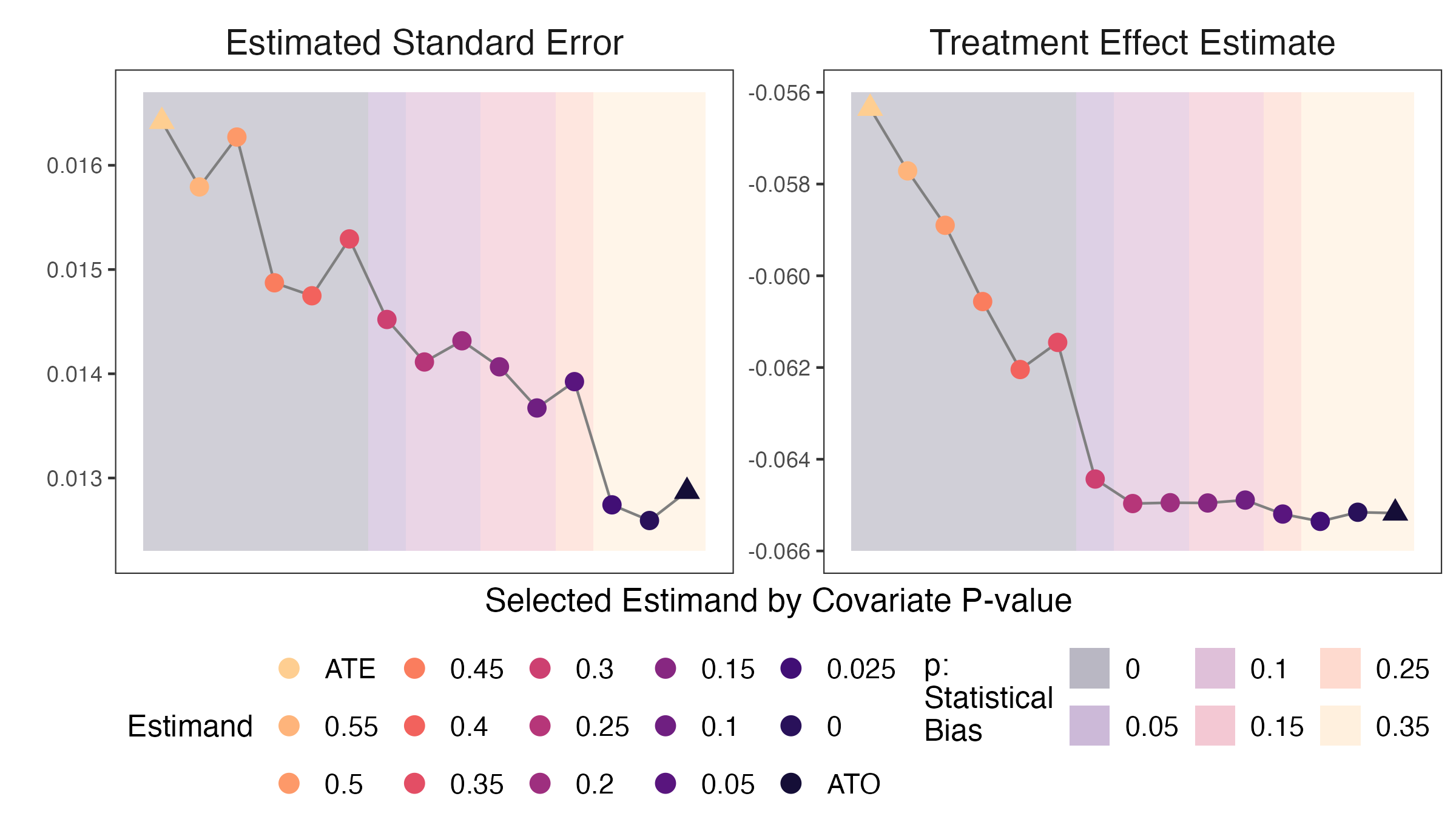}
    \caption{Characteristics of the ATE, %optimal 
    selected estimands, and ATO for the right heart catheterization study. The selected estimands are labeled by the lower bound of their corresponding estimand mismatch $p$-value contour. Generally, lighter colors indicate estimands with better performance with respect to one aspect of the bias decomposition; lighter points indicate smaller potential for estimand mismatch, while lighter background shading indicates smaller potential for statistical bias.
    }
    \label{fig:rhc_metrics_optimal}
\end{figure}

There are multiple ways to select a single %optimal 
estimand based on the tradeoffs between the three metrics. First, we reiterate that these estimands only differ when there is treatment effect heterogeneity. Therefore, in cases with minimal treatment effect heterogeneity, estimand mismatch may be a lower priority than statistical bias regardless of whether the ATE population is of interest. For this data, there is moderate to low evidence of treatment effect heterogeneity, which indicates we may want to prioritize minimizing estimator bias and standard error rather than estimand mismatch. When focusing only on these two metrics, the $p = 0$ estimand is the preferred estimand as it is in the largest statistical bias $p$-value contour and its estimator has the lowest estimated standard error (Figure \ref{fig:rhc_metrics_optimal}). If we want to maximize the performance of our selected estimand while targeting a population similar to the ATE population, we could choose either the $p = 0.05$ or $p = 0.10$ estimand as they, respectively, are in the highest statistical bias $p$-value contour or have the lowest estimated standard error of all estimands with estimand mismatch $p > 0.05$. If there is belief in or evidence of substantial treatment effect heterogeneity, we would potentially want to select an estimand with likely smaller estimand mismatch, such as the $p = 0.20$ estimand. For this analysis, any of the $p = 0$, $p = 0.05$, or $p = 0.10$ estimands may be ideal due to the limited treatment effect heterogeneity exhibited in the data.

%While in practice one would not look at the treatment effect estimates before selecting an estimand, 
For illustrative purposes, we examine the treatment effect estimates of the selected estimands (Figure \ref{fig:rhc_metrics_optimal}).
%we show the different treatment effect estimates for the set of %optimal 
%selected estimands in Figure $\ref{fig:rhc_metrics_optimal}$. 
The treatment effect estimate is around $-0.056$ for the ATE estimator and tends to decrease as the estimands become more similar to the ATO, whose estimate is around $-0.065$; all estimates indicate a negative effect of RHC on 30 day survival. The ATO and all estimands from estimand mismatch contours with $p < 0.35$ have similar treatment effect estimates around $-0.065$. 
%The ATE estimator has potential for substantial bias, as the statistical bias $p$-value is less than $0.05$ and further has large variance. 
A re-analysis of this data by \cite{hirano2001} using regression augmented estimators, which are more robust to model misspecification, found that when using propensity score and regression models with at least $15$ covariates, the treatment effect estimates ranged from $-0.059$ to $-0.067$. Since these results are similar to treatment effect estimates from the ATO and similar selected estimands, there is evidence that these estimands may yield estimates closer to the true ATE than the ATE estimator due to a reduction statistical bias while not incurring substantial estimand mismatch.

\section{Discussion}
\label{sec:discussion}
Covariate imbalance and a lack of overlap pose significant challenges when estimating a causal treatment effect in a target population.
To address these challenges, we have introduced a framework for characterizing and selecting causal estimands based on the population targeted and the corresponding estimator's statistical bias and variance. We have proposed a bias decomposition and design-based metrics that characterize bias due to estimand mismatch (i.e., deviation from the population of interest) and statistical bias, allowing one to characterize how the choice of estimand impacts performance along these facets. Our estimand selection procedure identifies a sequence of estimands whose corresponding estimators tend to have reduced MSE with respect the target estimand and a clear interpretation in respect to the bias-variance tradeoff. In addition, our selection procedure is flexible and allows the analyst to incorporate their domain-specific preferences. %for maintaining the target population versus reducing statistical bias and variance. 
%When applying our method to simulated data with a population target estimand of the ATE, the estimands selected through our procedure reduce estimator MSE by moderating the size of both bias and variance, rather than minimizing one of these attributes. 

While we present methods and results with respect to the ATE as the estimand of scientific interest, these methods can be applied to any other estimand such as the ATT or when transporting effects to a new population.  In addition, our methods could be extended to regression augmented estimators, though further work is needed to assess how well our proposed metric (Section \ref{sec:stat_bias_metric}) characterizes statistical bias in these cases. Furthermore, while we explore the set of estimands in Equation \eqref{eq:pot_estimands}, any potential set of estimands defined by weighted populations could be explored. %; for example, estimands with $c,d < 0$ or $c,d > 1$ could be of interest, especially when the target estimand is the ATT or ATC. 
Identifying a different set of potential estimands may be important when transporting effects as the there is no guarantee that a new population is similar to the weighted populations defined by Equation \eqref{eq:pot_estimands}.

We acknowledge that using our proposed procedure to identify an estimand could influence downstream inference. However, the majority of our procedure is design-based which helps protect against possible impacts on inference when using an estimand selected through this procedure. In fact, the procedure could be made entirely design-based. For example, one could assume a structure for the residual variance (e.g., homoscedasticity) such that some identifiable $h^*(\bm{x})$, that is a function of only covariates, achieves minimum asymptotic variance. Then, instead of selecting the $h_{c,d}(\bm{x})$ with the smallest estimated standard error for each estimand mismatch contour level, we could select the $h_{c,d}(\bm{x})$ that minimizes some distance to $h^*(\bm{x})$ (e.g., Euclidean distance on the $c,d \in [0,1] \times [0,1]$ space). Further, \cite{yang_asymptotic_2018} provide methods for incorporating uncertainty from both the design and analysis stage in the inferential procedure when using continuous weights that could be used. 
%  The \backmatter command formats the subsequent headings so that they
%  are in the journal style.  Please keep this command in your document
%  in this position, right after the final section of the main part of 
%  the paper and right before the Acknowledgements, Supporting Information (Web %  Materials),   and References sections. 
\section*{Funding}
The first author was supported by the National Science Foundation Graduate Research Fellowship Program under Grant No. 2237827.

\section*{Code and Data}

 The Git repository \url{https://github.com/m-barnard/estimand_selection} contains data and code to replicate the analyses and to implement the estimand selection procedure on a new dataset. 

%  This section is optional.  Here is where you will want to cite
%  grants, people who helped with the paper, etc.  But keep it short!

% \section*{Acknowledgements}

% The authors thank Professor A. Sen for some helpful suggestions,
% Dr C. R. Rangarajan for a critical reading of the original version of the
% paper, and an anonymous referee for very useful comments that improved
% the presentation of the paper.\vspace*{-8pt}

%  Here, we create the bibliographic entries manually, following the
%  journal style.  If you use this method or use natbib, PLEASE PAY
%  CAREFUL ATTENTION TO THE BIBLIOGRAPHIC STYLE IN A RECENT ISSUE OF
%  THE JOURNAL AND FOLLOW IT!  Failure to follow stylistic conventions
%  just lengthens the time spend copyediting your paper and hence its
%  position in the publication queue should it be accepted.

%  We greatly prefer that you incorporate the references for your
%  article into the body of the article as we have done here 
%  (you can use natbib or not as you choose) than use BiBTeX,
%  so that your article is self-contained in one file.
%  If you do use BiBTeX, please use the .bst file that comes with 
%  the distribution.  In this case, replace the thebibliography
%  environment below by 
%
\bibliographystyle{final_corrected_biom} 
\bibliography{refs.bib}

% \begin{thebibliography}{}

% \bibitem{ } Cox, D. R. (1972). Regression models and life tables (with
% discussion).  \textit{Journal of the Royal Statistical Society, Series B}
% \textbf{34,} 187--200.

% \bibitem{ }  Hastie, T., Tibshirani, R., and Friedman, J. (2001). \textit{The 
% Elements of Statistical Learning: Data Mining, Inference, and Prediction}.
% New York: Springer.

% \end{thebibliography}

%  If your paper refers to supporting web material, then you MUST
%  include this section!!  See Instructions for Authors at the journal
%  website http://www.biometrics.tibs.org

 \pagebreak

\center{\textbf{\Large Supplementary Materials}}

\justifying
\section*{Appendix A}
\raggedright
 \begin{proof}
    By Theorem 2 of \cite{li2018}, for a sample $\bm{X} = \{\bm{x}_1, \ldots, \bm{x}_n\}$, as $n \to \infty$,
\begin{equation*}
    nE_{\bm{x}}\{\Var(\hat{t}_{h}|\bm{X})\} \to \frac{1}{C_h^2}\int f(\bm{x})h(\bm{x})^2[v_1(\bm{x})/e(\bm{x}) + v_0(\bm{x})/\{1 - e(\bm{x})\}]d\bm{x}. 
\end{equation*}
where $v_z(\bm{x}) = Var\{Y(z)|\bm{X} = \bm{x}\}$ and $C_h = \int h(\bm{x})f(\bm{x})d\bm{x}$. Let $v \in \mathbb{R}^+$ and functions $h^*(\bm{x}),k_0(\bm{x}), k_1(\bm{x}) > 0$ for all $\bm{x}$.
%, and functions $k_0(\bm{x}), k_1(\bm{x}) > 0$ be well defined over the domain of $\bm{X}$. 
Now, consider
$  v_0(\bm{x}) = v\frac{k_0(\bm{x})\{1-e(\bm{x})\}}{h^*(\bm{x})\{k_0(\bm{x}) + k_1(\bm{x})\}}$ and $v_1(\bm{x}) =v\frac{k_1(\bm{x})e(\bm{x})}{h^*(\bm{x})\{k_0(\bm{x}) + k_1(\bm{x})\}}$  such that as $n \to \infty$,
\begin{align}
     nE_{\bm{x}}\{\Var(\hat{t}_{h}|\bm{X})\} &\to \frac{v}{C_h^2}\int f(\bm{x})h(\bm{x})^2\frac{k_0(\bm{x}) + k_1(\bm{x})}{h^*(\bm{x})\{k_0(\bm{x}) + k_1(\bm{x})\}}d\bm{x}, \\
     &\to \frac{v}{C_h^2}\int \frac{f(\bm{x})h(\bm{x})^2}{h^*(\bm{x})}d\bm{x}.
     \label{eq:simplified_asymp_var}
\end{align}
For clarity in notation, $E[\cdot]$ indicates $\int \cdot f(\bm{x})d\bm{x}$. Using the same argument as the proof of Corollary 1 in \cite{li2018}, by the Cauchy-Schwartz inequality
\begin{align*}
    E\{h(\bm{x})\}^2 &= \left[E\left\{\frac{h(\bm{x})}{\sqrt{h^*(\bm{x})}}\sqrt{h^*(\bm{x})}\right\}\right]^2, \\
    &\leq E\left\{\frac{h(\bm{x})^2}{h^*(\bm{x})}\right\}E\{h^*(\bm{x})\},
\end{align*}
where equality holds when $h(\bm{x}) \propto h^*(\bm{x})$. Applying this result to the right-hand side of \eqref{eq:simplified_asymp_var}, we have that as $n \to \infty$
\begin{equation*}
    n\min_h(E_{\bm{x}}\{\Var(\hat{t}_{h}|\bm{X})\}) \to \frac{v}{C_h^2}\int f(\bm{x})h^*(\bm{x})d\bm{x},
\end{equation*}
and $h(\bm{x}) \propto h^*(\bm{x})$ gives the smallest asymptotic variance for $\hat{\tau}_h$ among all $h$.
\end{proof}

\pagebreak
\setcounter{figure}{0} 
\section*{Supplementary Tables and Figures}

\begin{table}[h!]
\caption{ Common causal estimands and their corresponding $h(\bm{x})$, balancing weights 
\citep{li2018}, and population.}
\centering
\begin{tabular}{cccc} \hline
Estimand &  $h(\bm{x})$    &  $w_0(\bm{x}), w_1(\bm{x})$     &  Target Population \\ \hline
ATE          & $1$     & $1/\{1-e(\bm{x})\}, 1/e(\bm{x})$       & Combined      \\
ATT           & $e(\bm{x})$  & $e(\bm{x})/\{1-e(\bm{x})\}, 1$      & Treated      \\
ATC         & $1-e(\bm{x}) $   & $1, e(\bm{x})/\{1-e(\bm{x})\}$   & Control      \\
ATO          & $e(\bm{x})\{1-e(\bm{x})\}$& $e(\bm{x}), 1-e(\bm{x})$ & Overlap      \\   \hline    
\end{tabular}
\label{tab:common_estimands}
\end{table}
\pagebreak

    \begin{figure}[h!]
    \centering
    \hspace*{-0.3in}
    \includegraphics[scale = 0.78]{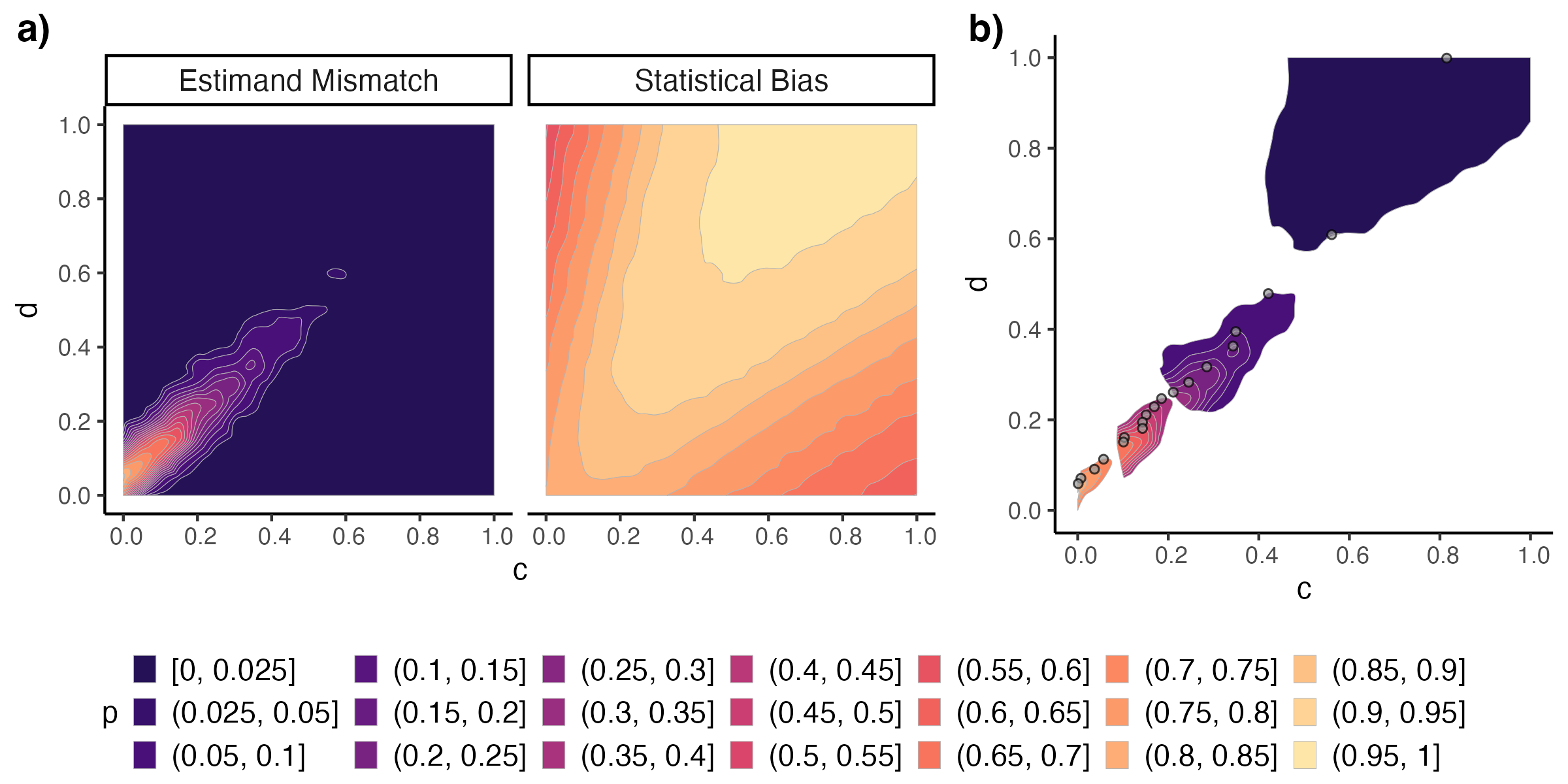}
    \caption{Subfigure a) shows the contours of the $p$-values corresponding to estimand mismatch and statistical bias. Data is interpolated using cubic splines in order to smooth contours. Subfigure b) shows the intersection of each estimand mismatch $p$-value contour with the largest statistical bias $p$-value contour level (color indicates the estimand mismatch $p$-value contour level). The points in subfigure b) show the %optimal 
    selected estimands for each estimand mismatch $p$-value contour.}
    \label{fig:example_contour}
    \end{figure}

    \begin{figure}[h!]
         %\hspace*{-5em}
        \begin{subfigure}{\textwidth}
          \centering
          \includegraphics[draft = false, width=.8\linewidth]{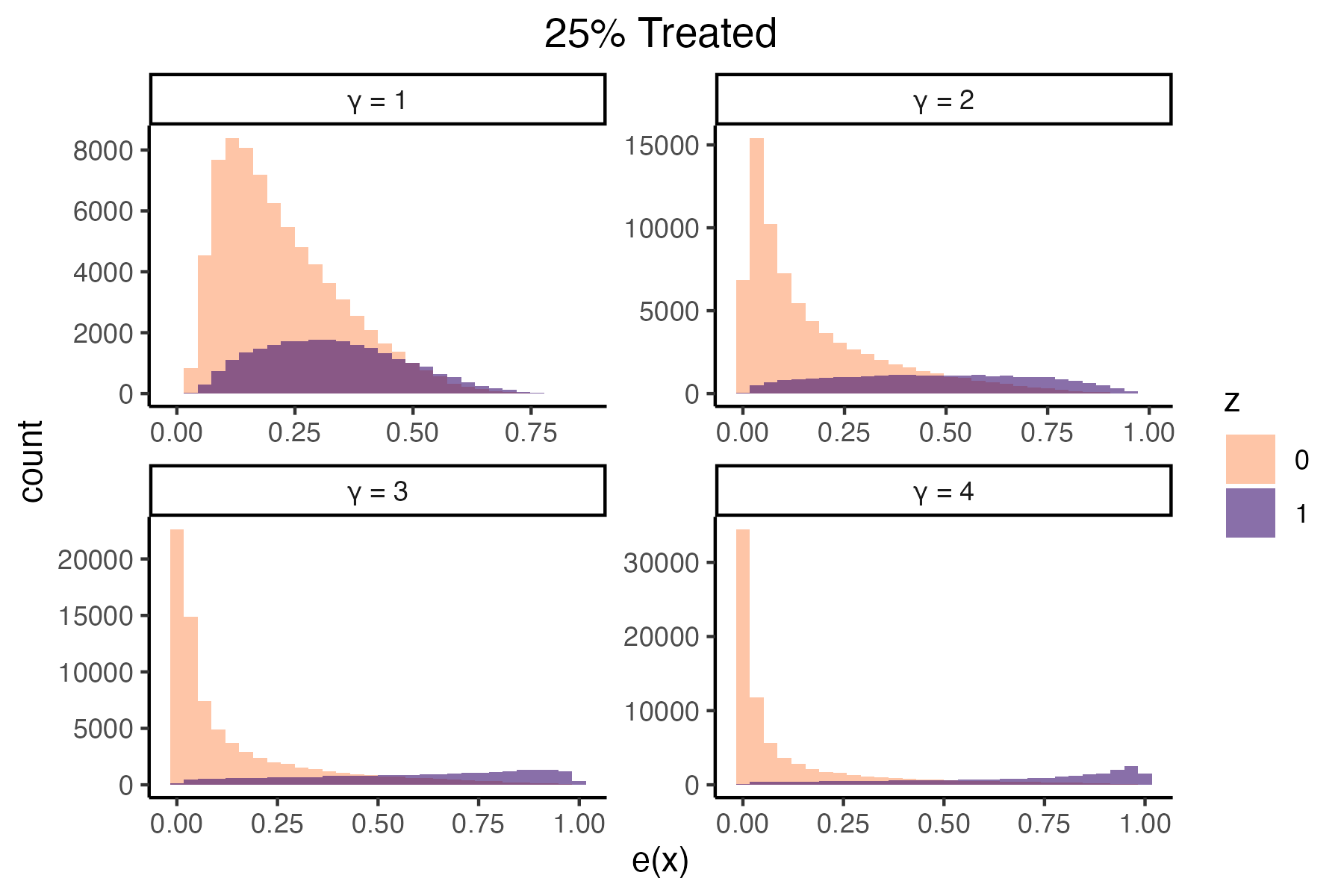}
          %\caption{}
          \label{fig:step2}
        \end{subfigure}%\hspace*{-5em}
        \\
        \begin{subfigure}{\textwidth}
          \centering
          \includegraphics[draft = false, width=.8\linewidth]{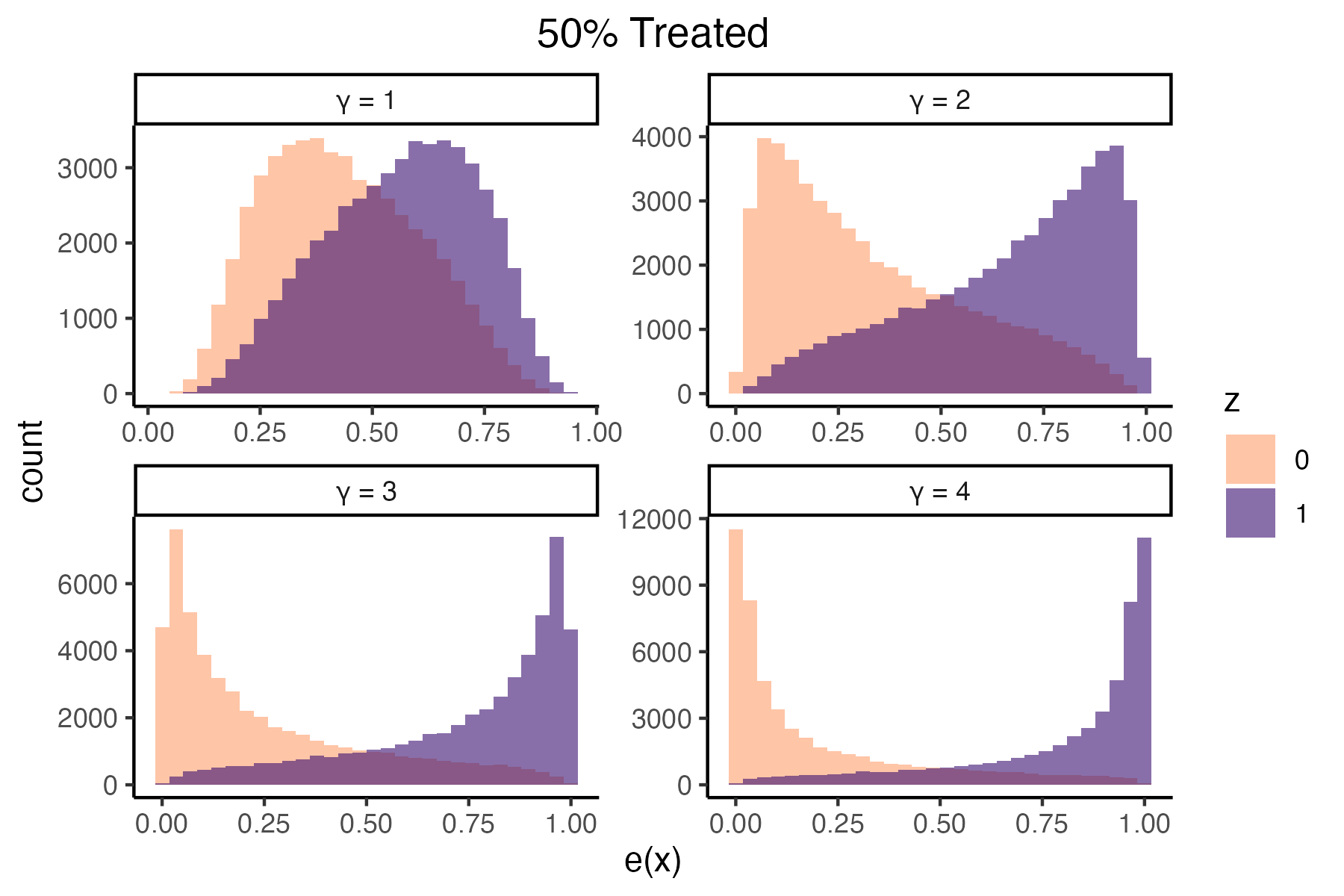}
          %\caption{}
          \label{fig:step3}
        \end{subfigure}
    \caption{ Propensity score overlap for all simulation scenarios. %Panel a) and b) show overlap for simulated data with 25\%  and 50\% of individuals treated, respectively. 
    The top and bottom four panels show overlap for simulated data with 25\%  and 50\% of individuals treated, respectively. Note that $\gamma$ indicates the level of propensity score overlap where $\gamma = 1$ is high overlap and $\gamma = 4$ is low overlap.}
    \label{fig:ps_overlap}
    \end{figure}

\begin{figure}[h!]
    \centering
     \hspace*{-0.3in}
    \includegraphics[draft = false, scale = 0.8]{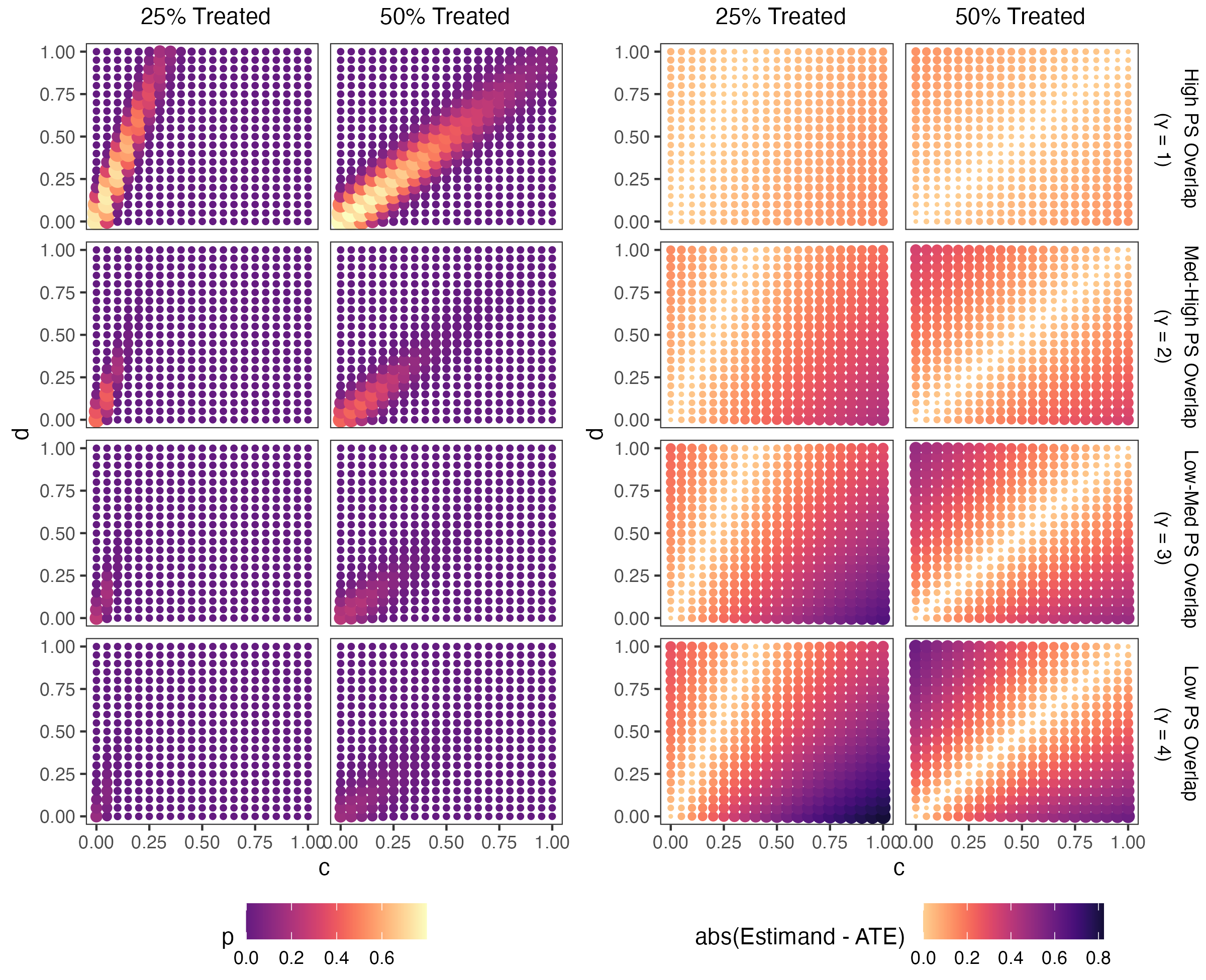}
    \caption{ The left panel presents the average of the minimum of the two estimand mismatch related $p$-values, calculated as described in Section \ref{sec:mismatch_metric}, for each estimand. The right panel presents the absolute difference between the true estimands and the true ATE (i.e., the absolute estimand mismatch) under the medium $\Delta$ heterogeneity scenario when simulating the treatment effect as a linear function of the propensity score. Columns distinguish the 25\% and 50\% treated scenarios while rows distinguish the different propensity score overlap scenarios.}
    \label{fig:p_cov_supp}
\end{figure}

\begin{figure}[h!]
    \centering
    \includegraphics[draft = false, scale = 0.8]{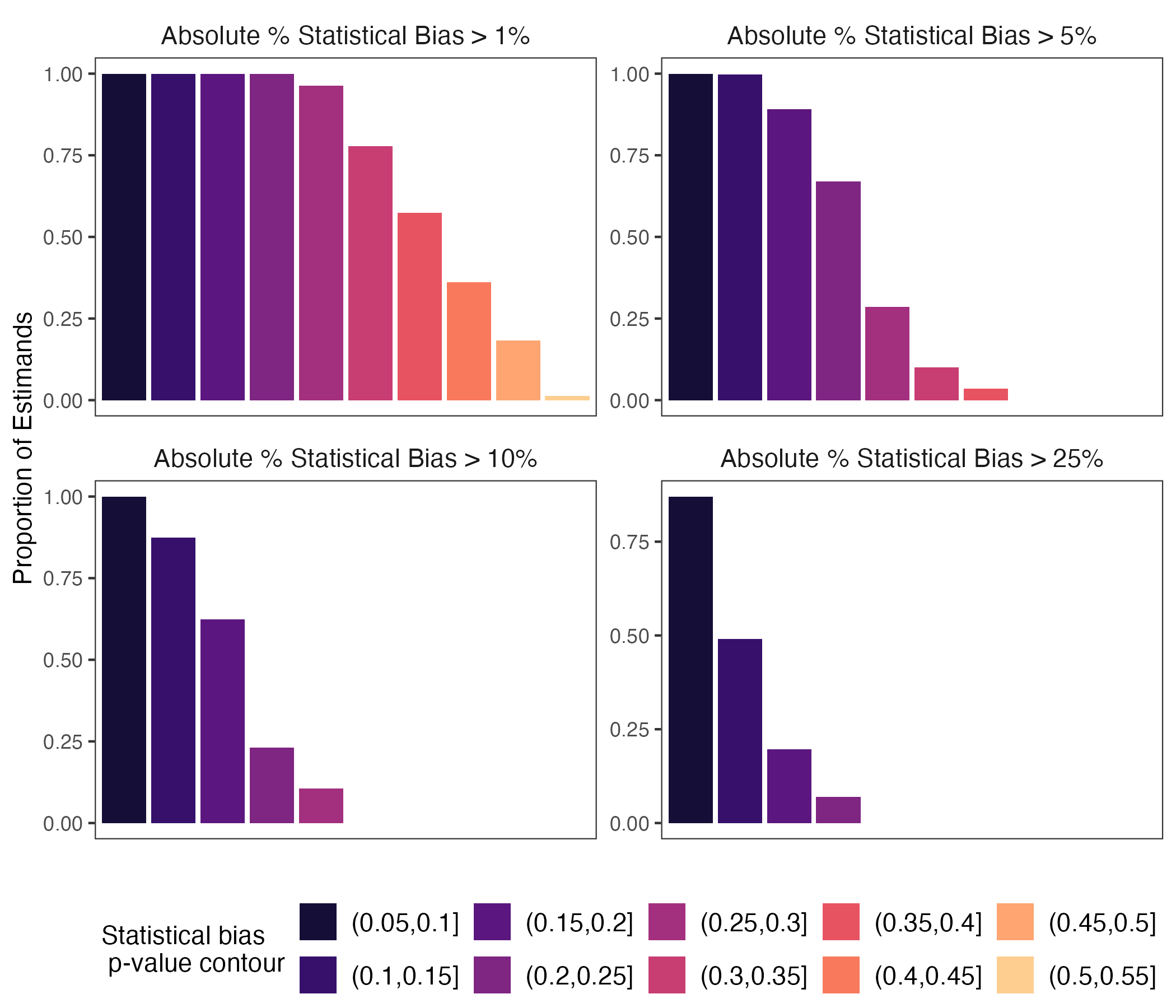}
    \caption{ Proportion of estimands within each statistical bias associated $p$-value contour with absolute percent bias greater than 1\%, 5\%, 10\%, and 25\%.}
    \label{fig:p_ps_supp}
\end{figure}

% \begin{figure}[h!]
%     \centering
%     \includegraphics[draft = false, scale = 0.8]{figs/se.png}
%     \caption{Average bootstrapped standard error for all estimators corresponding to the estimands defined by Equation \eqref{eq:pot_estimands} across all simulation scenarios. Each of the 16 subplots corresponds to a single simulation scenario as labeled by the top and right axes.}
%     \label{fig:se}
% \end{figure}

% \begin{figure}[h!]
%     \centering
%     \includegraphics[draft = false, scale = 0.8]{figs/se_emp_sd.png}
%     \caption{Spearman correlation between the true estimator standard error (as measured by the empirical standard deviation) and the average bootstrapped standard error across all simulation scenarios. Each of the 16 subplots corresponds to a single simulation scenario as labeled by the top and right axes. }
%     \label{fig:se_emp_sd}
% \end{figure}

\begin{figure}[h!]
    \centering
    \hspace*{-0.5in}
\includegraphics[draft = false, scale = 0.7]{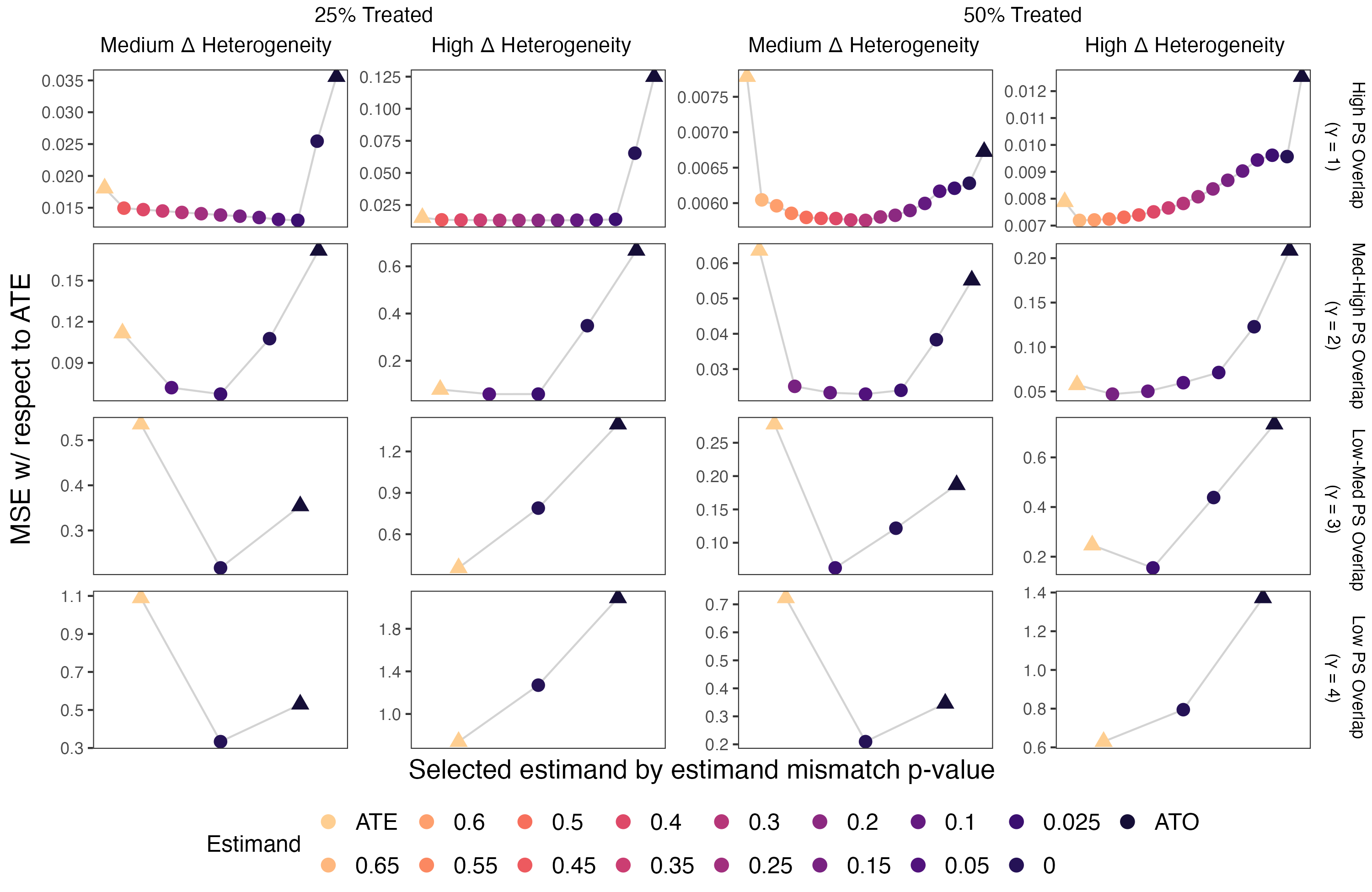}
    \caption{ Mean squared error (MSE) with respect to the true ATE for the estimators for the ATE, the estimands selected with our procedure, and the ATO. Each of the 16 subplots corresponds to a single simulation scenario as labeled by the top and right axes. Estimands are labeled by the lower bound of their corresponding estimand mismatch $p$-value contour. Estimands selected (i.e., estimand mismatch $p$-value contours present) in at least 900 of the simulated datasets are shown.}
    \label{fig:mse_900n}
\end{figure}

\begin{figure}[h!]
    \centering
    \hspace*{-0.4in}
\includegraphics[draft = false, scale = 0.8]{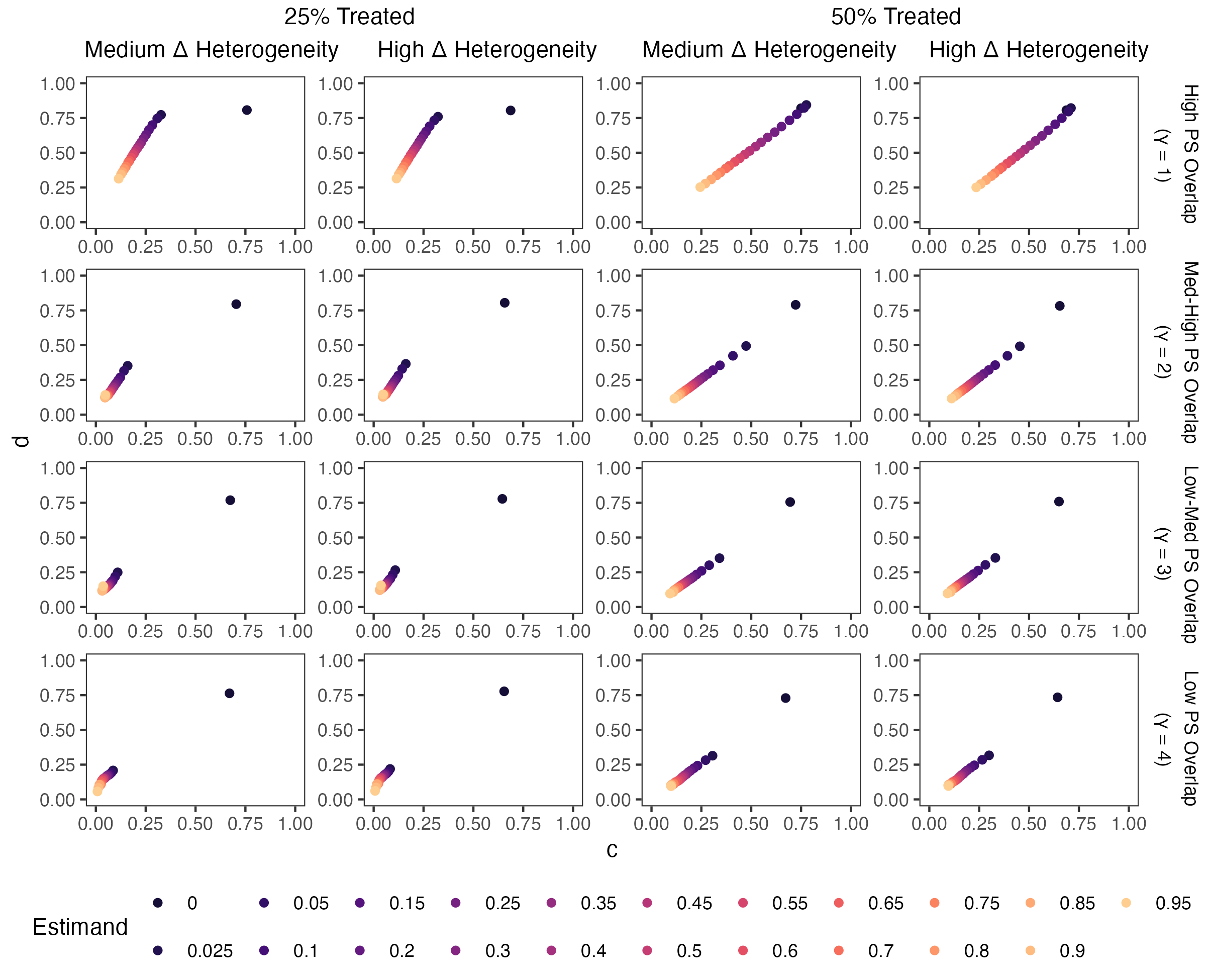}
    \caption{ Average of the selected estimands (i.e., average of $c$ and $d$) for each estimand mismatch $p$-value contour across all possible simulated datasets. Each of the 16 subplots corresponds to a single simulation scenario as labeled by the top and right axes. Estimands are labeled by the lower bound of their corresponding estimand mismatch $p$-value contour.}
    \label{fig:avg_est}
\end{figure}

\begin{figure}[h!]
    \centering
\includegraphics[draft = false, scale = 0.9]{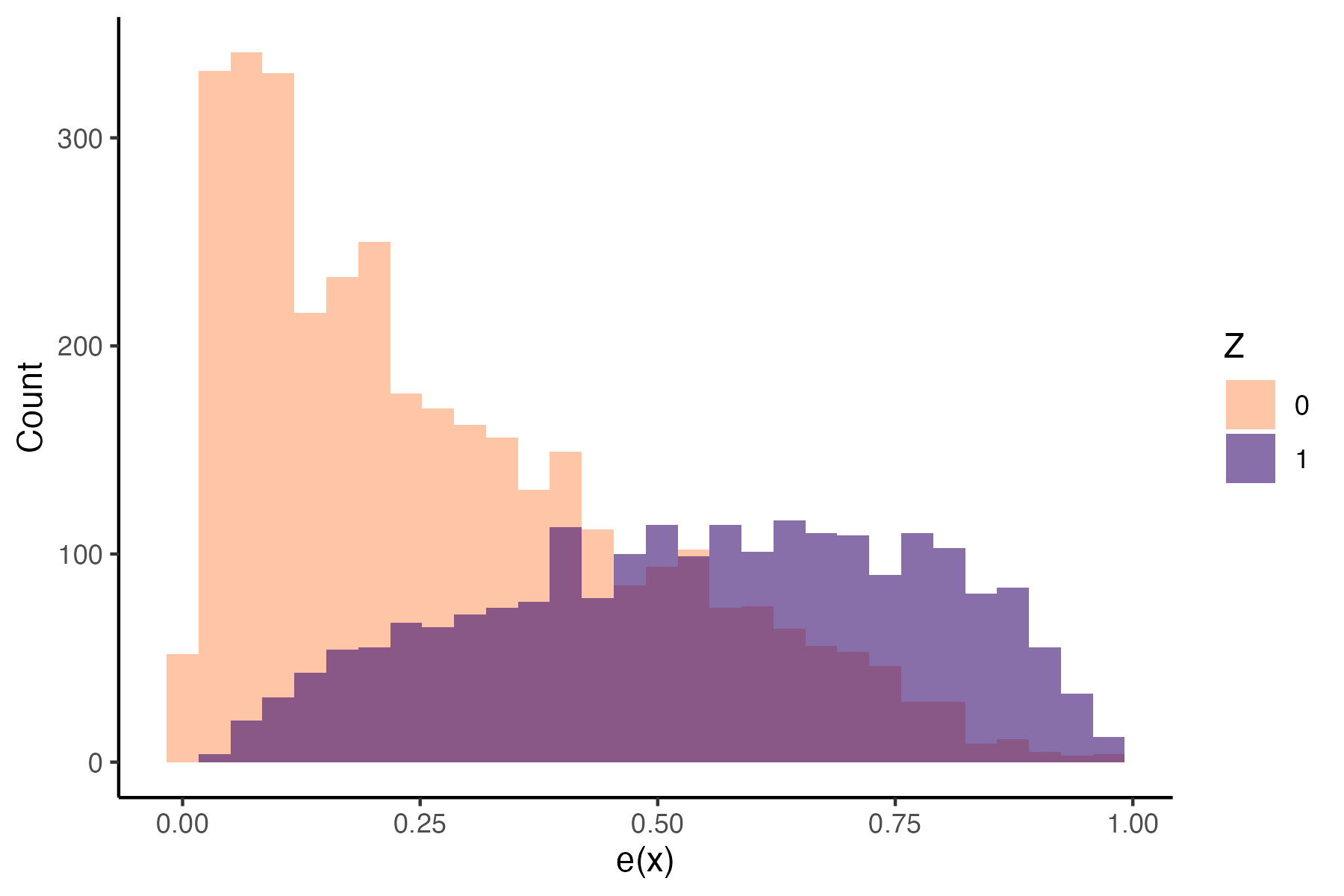}
    \caption{Estimated propensity score distributions of treated and control groups for the right heart catheterization study \citep{connors1996}.}
    \label{fig:rhc_ps_dist}
\end{figure}

\begin{figure}[h!]
    \centering
\includegraphics[draft = false, scale = 0.9]{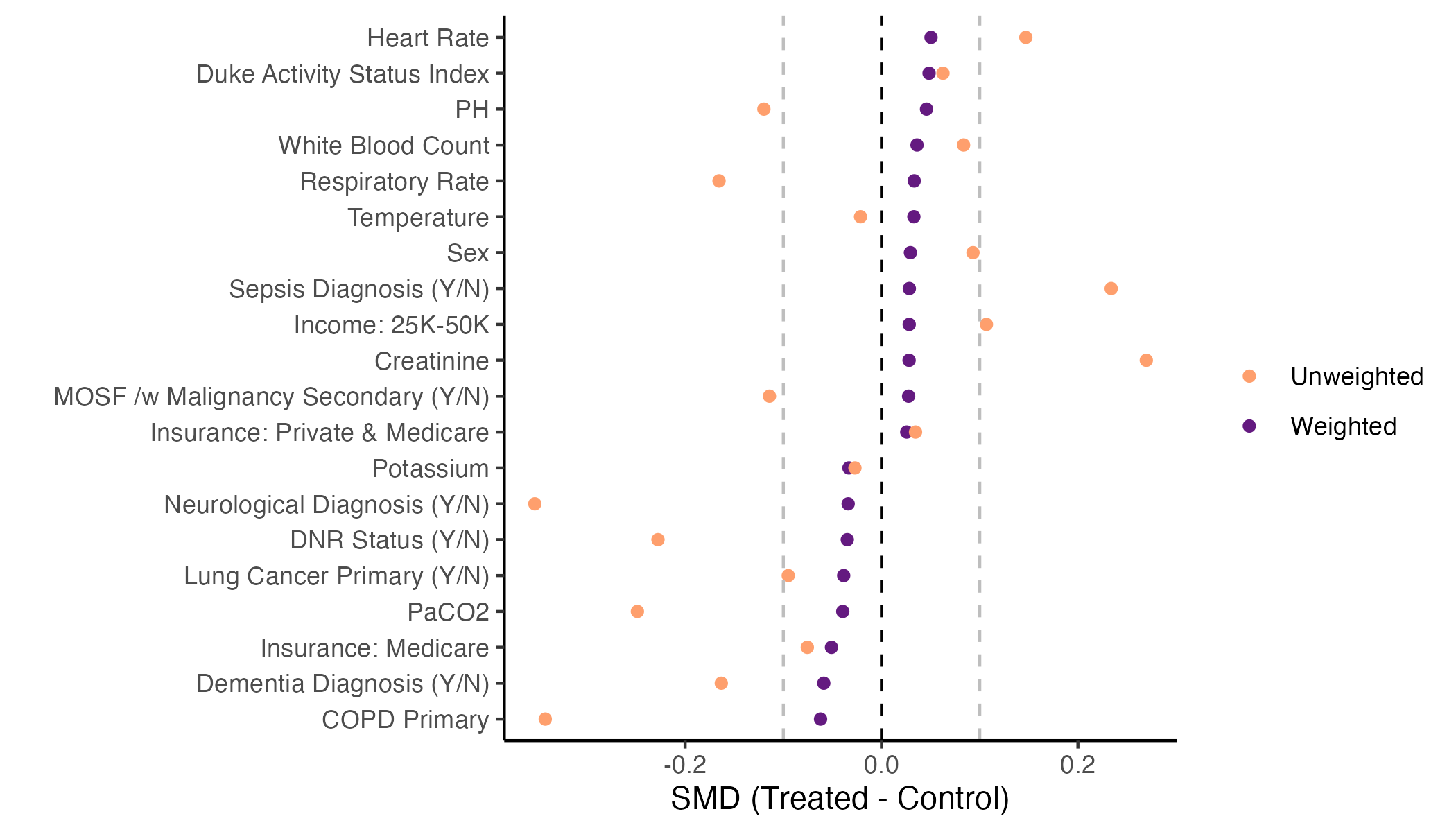}
    \caption{Standardized mean difference (SMD) of the twenty most imbalanced covariates after IPW weighting for the right heart catheterization study \citep{connors1996}.}
    \label{fig:rhc_cov_balance}
\end{figure}

\end{document}